\magnification=\magstep1
\def\reidelbaselines{\baselineskip=10.10pt
                     \lineskip=0pt
                     \lineskiplimit=0pt}
\def\oneskip{\vskip\baselineskip}
\def\doublespace{\baselineskip=24pt\parskip=0pt plus 1 pt}
\hoffset=0.5 truecm
\voffset=1.0 truecm
\hsize=16.0 truecm
\vsize=22.5 truecm
\headline {\hss\tenrm\folio\hss}
\nopagenumbers
\doublespace
\reidelbaselines
\centerline{\null}
\noindent
{\bf Starcounts redivivus II:
Deep starcounts with Keck and HST 
and the luminosity function of the Galactic halo}
\oneskip
\parindent=45pt
\parskip 0pt
\oneskip
\item\item{} Iain \ Neill \ Reid
\footnote{$^1$}
{Visiting Research Associate, The Observatories of the Carnegie 
Institution of Washington}
\item\item{} 105-24
\item\item{} California Institute of Technology
\item\item{} Pasadena, CA 91125, USA
\oneskip
\item\item{} Lin \ Yan
\item\item{} European Southern Observatory
\item\item{} Karl-Schwarzschild-Strasse 2
\item\item{} 85748 Garching bei Munchen, Germany
\oneskip
\item\item{} Steve \ Majewski$^1$
\item\item{} University of Virginia
\item\item{} Department of Astronomy
\item\item{} P.O. Box 3818
\item\item{} Charlottesville, VA 22903-0818, USA
\oneskip
\item\item{} Ian \ Thompson
\item\item{} Observatories of Carnegie Institution of Washington
\item\item{} 813 Santa Barbara St.
\item\item{} Pasadena, CA 91101-1292
\oneskip
\item\item{} Ian \  Smail$^1$
\item\item{} University of Durham
\item\item{} Department of Physics
\item\item{} Durham, DH1 3LE, England
\oneskip
\parindent=1.0 truecm
\vskip 0.5 true in
Submitted to Astronomical Journal, 30 April 1996; revised, July 1 1996
\vskip 1.2 true in
\noindent
Based partly on observations obtained at the W.M. Keck Observatory, which
is operated by the Californian Association for Research in Astronomy
for the California Institute of Technology and the University of California,
and partly on observations obtained with the  NASA/ESA Hubble Space Telescope,
obtained at the Space Telescope Science Institute, which is operated by 
AURA, Inc., under NASA contract NAS5-26555
\vfill\break
\oneskip
\noindent
{\bf Abstract }
\oneskip
\noindent
We have combined deep starcount data with Galaxy model predictions to 
investigate how effectively such measurements probe the faint end of
the halo luminosity function. We have tested a number of star/galaxy
classification techniques using images taken in 0.5 arcsecond seeing 
with LRIS on the Keck telescope, and we find that different combinations
of these techniques can produce variations of 10 \% in the inferred
starcounts at R=22.5 and 30 \% at R=24.5 magnitudes. The decreasing 
average angular size of galaxies with fainter magnitude effectively 
limits ground-based work to R $<$ 25.5 magnitudes. The higher angular
resolution provided by HST allows one to probe at least 2 magnitudes
fainter, but the small field-size is a significant limitation. In
either case, our models show that the contribution from halo subdwarfs
is effectively limited to colours of (R-I) $<$ 1.0, with the redder stars
being members of the Galactic disk. The apparent increase in number
density for M$_V > 10$ in the derived luminosity function is a result of
contributions from disk stars at fainter absolute magnitudes and
does not provide evidence for an
upturn in the halo subdwarf mass function. Indeed, starcount data alone
are {\sl not} an
effective method of probing the shape of the halo luminosity function
close to the hydrogen-burning limit. Finally, we examine
how the Hubble Deep Field observations can be used to constrain the 
contribution of various stellar components to the dark-matter halo.
\oneskip
\noindent
{\bf 1. Introduction }
\oneskip
\noindent
Over the past decade considerable effort, both observational
and theoretical, has been directed towards deriving a more accurate 
determination of the stellar luminosity function of the Galactic disk, 
with particular emphasis on the behaviour at low luminosities. Most
recent derivations (Tinney, 1993; Kirkpatrick et al, 1994; Reid, Hawley
\& Gizis, 1995) show a broad peak at M$_V \sim$ 12 (or equivalent) and
subsequent decline, with the hint of an upturn at the faintest magnitudes.
Converting to a mass function is more controversial, largely reflecting
the poor observational constraints on the (disk) mass-luminosity relation at
low masses, but the majority of studies show a turnover well before
the hydrogen-burning limit, while even the most optimistic analyses
(Kroupa, 1995) find a mass function increasing no faster than 
$\Psi(M) \propto m^{-1}$ for $M < 0.5 M_\odot$. In either case, it is 
clear that low-mass stars ($M < 0.2 M_\odot$) and brown dwarfs make only 
a minor contribution to the total mass of the Galactic disk.
\oneskip
The situation is less well defined for the Galactic halo, primarily 
because of the greater difficulties involved in identifying a pure sample of 
halo stars.
\footnote {$^2$}
{Recent years have seen some confusion caused by the use of the term `halo'
to refer to the dark matter whose presence is indicated by rotation
curve analyses. In this paper we adopt the conventions of the 1957
Vatican conference (O'Connell, 1958)
and use the term to refer to the old stellar population
that represents the field counterpart of the globular cluster system.
All discussions of the dark-matter halo are specifically identified as such.} 
Observational efforts have followed two main paths: studies of globular
cluster members and surveys of field halo subdwarfs. The initial results from
the former approach (Richer et al, 1991; Drukier et al, 1993) suggested 
strongly a luminosity
function that continued to increase steeply towards lower luminosities, with
no suggestion of a turnover. However, more recent, higher spatial-resolution
HST observations (Paresce et al, 1995; Elson et al, 1995; King \& Anderson, 1996) 
contradict this result, finding a broad maximum at $M_V \sim +11$. HST
observations are required of clusters such as NGC 6752, where the ground-based 
data suggest steeply increasing luminosity functions (Richer et al, 1991)
before this issue can be resolved.
\oneskip
Given the likelihood of internal dynamical evolution, as well as possible
selection effects in the cluster disruption process, globular clusters cannot 
necessarily be taken as representative of the halo at large, but studies of
the field have yielded the same dichotomy of results. Dahn et al (1994 - DLHG),
studying a sample of nearby (r $\le$ 300 pc.) high proper-motion stars,
have derived a luminosity function similar in form to the Paresce et al
NGC 6397 function. On the other hand, Richer and Fahlman (1992) have used
deep CCD starcounts to attempt to
isolate a sample of stars at large distances above the
disk and, based on that sample, have argued that the halo luminosity function
increases steeply and continuously to at least $M_V \sim +14$ - close to the
H-burning limit for metal-poor halo subdwarfs. 
There are possible shortcomings associated with either of these approaches:
the proper motion stars are selected from the Luyten
LHS catalogue (Luyten, 1979), and it is possible the results could be biased
through kinematically-related selection effects. Alternatively, the deep 
starcount analysis may suffer from errors introduced either through
uncertainties in star/galaxy separation or through misclassification of
disk stars as halo subdwarfs. In this paper we address the latter issues,
combining both ground-based starcount data from the Keck 10-metre 
telescope and fainter data from the Hubble Deep Field with predictions
derived from improved starcount models. Section 2 discusses the uncertainty
inherent in applying various star/galaxy separation techniques to
good-seeing ground-based data; in section 3 we consider the expected
colour-magnitude distribution of the different stellar populations,
with particular emphasis on the redder, lower-mass stars; section 4 compares
our deep starcounts with the predictions of several models; and section 5
presents our conclusions. 
\oneskip
\noindent
{\bf 2. Star-Galaxy separation at faint magnitudes}
\oneskip
\noindent
Accurate star-galaxy separation is essential at faint magnitudes. Reid 
\& Majewski (1993 - RM93) discussed this problem extensively in the context 
of deep (B$_J \sim 22$) photographic data towards the NGP. Figure 1
shows a more general illustration of the magnitude of the problem, 
comparing the stellar and galaxian distributions in fields at
relatively high galactic latitude (b=45$^o$, 60$^o$ and 90$^o$). 
The galaxy counts are
taken from Metcalfe et al. (1995) and Smail et al (1995: S95).
The star counts are based on one of the models described 
in the following section, but
the total counts are relatively insensitive ($< 10 \%$) to the choice of 
parameters. It is clear that the surface density of stars equals that of 
galaxies at relatively bright magnitudes - B $\sim$ 21, R $\sim$ 18.8 and 
I $\sim$ 18.5. Moving to
fainter magnitudes, galaxies increase in number by a factor of at least 2 per
magnitude (S95) while the slope of the stellar counts flattens significantly
as one moves to larger distances and runs out of Galaxy.
Thus, by $R \sim 24$ galaxies outnumber stars by a factor
of $\sim$40:1, and a small error in galaxy classification becomes a 
substantial uncertainty in the starcounts. 
\oneskip
Most galaxies have colours of (V-R) $\sim 0.5 \pm 0.25$. Hence, while
one can expect contamination by only a few (unusual) galaxies in a
sample of red, low-mass star candidates, there is substantial overlap
between the galaxies and the locus of halo subdwarfs 
in both the colour-magnitude and colour-colour plane. We have illustrated
this in figure 2. Figure 2a shows the two-colour (V-R)/(R-I) relation
defined by nearby disk stars (Bessell, 1991) and halo subdwarfs (photometry
by Bessell, priv. comm.). The latter stars follow a similar two-colour
relation, lying at most 0.15 mag. to the blue in (R-I) 
at a given (V-R) colour. In constructing their faint stellar sample, 
Richer \& Fahlman (1992) applied both image structure and photometric criteria, 
rejecting objects lying more than 2$\sigma$ from the
Population I/II (disk/halo) stellar locus. We can estimate the
effectiveness of that technique by applying the same criterion to our
sample.
\oneskip
Figure 2b plots the two-colour diagram for galaxies brighter than R=25 in 
the PSR 1640 field, where we have superimposed the disk stellar sequence. 
The solid lines outline the locus corresponding to 
2$\sigma$ uncertainties of $\pm 0.14$ magnitudes in either colour -
matching the formal uncertainties for R $\sim 23.5 \pm 0.5$. Of the
204 galaxies in this magnitude range, 138 (67 \%) lie within 2$\sigma$
of the stellar sequence. The formal uncertainties at R$\sim 23$ are lower,
$\sim 0.035$ magnitudes in each colour, but 115 of 158 galaxies with
23 $\le R \le$ 24 mag. fall within the $\pm 2 \sigma$ stellar locus.
Thus, while VRI colours are an efficient means of identifying some quasars
and low-redshift AGNs, we conclude that if used alone they provide only weak 
star/galaxy discrimination.
\oneskip
Overall, morphology remains the most effective means of discriminating
between stars and galaxies. However, S95 have shown that the
median galaxian half-light radius is $\sim$ 0.4 arcseconds at R=24
and declines to 0.2 arcseconds by R=25.5 (S95, figure 4). As a result,
a significant fraction of galaxies with R $>$ 25.5 
are indistinguishable from point sources in ground-based data, and
any selection criterion based on
morphology cannot be expected to be successful at such faint magnitudes.
\oneskip
We have undertaken an investigation of the accuracy of star/galaxy
classification at brighter magnitudes using images taken in excellent
seeing with the Low Resolution Imaging Spectrograph (LRIS, 
Oke et al, 1995) on the Keck telescope.
The observations cover two fields at intermediate galactic latitude,
(l=41$^o$, b=38$^o$ - PSR 1640) and (l=88$^o$, b=-26$^o$ - PSR 2229), with
LRIS providing a pixel scale of 0$"$.22 and a field of view of 5'.7 by
7'.3. Full details of the data reduction are given elsewhere (S95).
The exposure times and FWHM are listed in
Table 1. The V-band exposure time is relatively short in field PSR2229, 
we have not used these observations in our analysis. All other
observations have similar limiting magnitudes as
the Richer and Fahlman (1992 - RF) CFHT observations. Photometric calibration
is based on observations of Landolt (1992) standards, and we have adopted
line-of-sight reddening values of E$_{B-V}$=0.07 mag. in both fields
(Stark et al, 1992). Note that while the PSR 1640 field is in the first
quadrant, the closest that that line-of-sight approaches the Galactic
Centre is 6.4 kpc (at a heliocentric distance of 4.8 kpc and at 3 kpc.
above the Galactic Plane). Thus possible contamination by bulge stars is not a 
problem.
\oneskip
Given the fundamental limit imposed on star/galaxy separation by the
decreasing average half-light radius of field galaxies (S95), we have
restricted our sample to sources brighter than R= 25 magnitude. The
formal photometric uncertainties in field PSR1640 are 0.08 mag. at R=25 and V=25,
and 0.15 mag. at I=25 magnitude, but the uncertainty in V rises to
$\sim 0.2$ magnitudes at V=26. Thus, a typical halo-star at R=25
has uncertainties of $\sim 0.22$ magnitudes in (V-R) and $\sim 0.13$
magnitudes in (R-I). In field PSR2229 the uncertainties are slightly
higher, with $\sigma_R = 0.1$ mag. at R=25 mag. and $\sigma_I = 0.23$ mag.
at I=25 magnitude.
\oneskip
We have applied a number of star/galaxy separation techniques to these
data and compared the results in an attempt to determine their relative
utility and reliability. 
In both fields the R-band data were taken in the best seeing 
conditions, and we have used these frames as the primary source of
our morphological classification. These data are sufficiently deep that
the limiting factor for star/galaxy separation is the spatial resolution (seeing)
rather than either photon statistics or incompleteness. 
\oneskip
In addition to analysing the real stellar images, we have also generated 200
artificial stars (using the point spread function given by DAOPHOT) in
each field, distributing their magnitudes between R=20 and R=25. These
stars have been placed at carefully chosen, relatively clear locations in 
each frame, with the appropriate photon- and sky-noise added. We have 
analysed these fake objects using the same techniques that are applied to
the real data, and use the results to provide an indication of both
the stellar locii and the dispersions as a function of magnitude. 
This analysis does not allow for the additional scatter caused by merged 
and overlapping images, but the source density is low enough that these
problems are relatively insignificant ($< 5 \%$) even at R=25 magnitude.
The results of the simulations are plotted, with the real 
data, in figure 3-6. 
\oneskip
We have used
the following morphological parameters to define stellar samples:
\oneskip
1) image ellipticity, defined as e = (a/b) - 1.0, where b and a are the 
semi-major and semi-minor axes of the best-fit isophotes as determined
by the profile-fitting routines of the SExtractor image analysis package
(Bertin, 1995). S95 describe the package in more detail. The resulting
ellipticity distribution as a function of magnitude is plotted in
figure 3a for PSR 1640 and figure 3b for PSR 2229. Figure 3c shows the 
distribution of the artificial stars and, based on that, we have
rejected all objects with ellipticity greater than 0.3 from the sample
as galaxies.
\oneskip
2) the $\chi$ parameter from DAOPHOT (Stetson, 1987). We defined the
reference point-spread function (PSF) using twelve isolated,
bright (and unsaturated) stars in each field, and the $\chi$ parameter
describes the goodness of fit between each image and the reference template.
The optical design of LRIS leads to small variations in the PSF across
the field, and ideally, one would prefer a larger number of reference stars, 
but crowding in these intermediate latitude fields limits the number of 
calibrators. While our tests show that DAOPHOT takes the psf variations 
successfully into account, systematic effects may make a small additional
contribution to the observed scatter.
Figures 4a and 4b plot this parameter against magnitude for the two fields,
with figure 4c showing the artificial star distribution.
All objects with $\chi > 3.0$ are rejected as galaxies.
\oneskip
3) log (I$_{peak}$ - I$_{sky}$)/I$_{sky}$ {\sl vs.} isophotal magnitude - 
i.e., a measure of the
compactness of the image profile. This is the parameter defined by Jones
et al (1991) and used (with methods 1 and 2) by RF as their primary 
star/galaxy discriminant. (They imposed the additional criterion of
requiring that stars lay within 2$\sigma$ of the VRI stellar locus.)
Data for objects in our two fields are plotted 
in figures 5a and 5c (note that stars saturate at R $<$ 20.5 magnitudes).
We have computed the mean
relations defined by the artificial stars (plotted in figures 5b and 5d
respectively) and the lines delineate the $\pm 5 \sigma$ envelope
about those relations.
\oneskip
4) I$_{peak}$/$\Phi(<0".6)$, where $\Phi(<0".6)$ is the total flux within
a 0$"$.6 diameter aperture centred on the source. Again, this is a measure of
the compactness of the stellar profile and the results are shown in
figures 6a and 6c, where the $\pm 5 \sigma$ envelope defined by the
artificial star samples (figures 6b and 6d) are also plotted.
\oneskip
Table 2 lists the relative numbers of stars and 
galaxies as a function of magnitude if we apply various combinations of
these criteria to our data. Applying the first two criteria gives only
a crude star/galaxy separation, with only the obviously-extended galaxies
being removed from the sample. Methods 3 and 4 are obviously more stringent,
even when using a $\pm 3 \sigma$ cut rather than the $\pm 5 \sigma$ limits
plotted in figures 5 and 6. Table 2 shows the results both  of
applying these two methods separately (but in conjunction with methods 1 and 2)
and of demanding that `stars' are objects which meet all four criteria. 
As one might expect, the last set of requirements (1+2+3+4) gives the
smallest sample, with number counts that are lower than those derived
based on the RF criteria (1+2+3) by $\sim 10 \% $ at R=22.5, $\sim 20 \%$
at R=23.5 and $\sim 30 \%$ at R=24.5 magnitude. Applying additional
criteria can only reduce the size of the final sample, and it is likely
that the (1+2+3+4) sample is over-conservative in definition. Nonetheless,
we conclude that, at least for our Keck ground-based data, uncertainties 
in the morphological classification imply uncertainties in the final 
starcounts of $\pm 10 \%$ at R=23.5 and $\pm 15 \%$ at R=24.5. The
addition of colour criteria should reduce these uncertainties by 
approximately two-thirds - note the excellent agreement in total
numbercounts between the RF observations and the predictions of the 
starcount models described in section 4.1.
\oneskip
\noindent
{\bf 3. Faint red stars and the halo}
\oneskip
\noindent
Given a reliable faint stellar sample, the next step is segregating sheep
and goats - sorting which stars are members of the disk and which are
members of the halo population. The
first point that should be made is that, despite recent analyses
(Bahcall et al, 1994; Santiago, Gilmore \& Elson, 1996), the lowest-luminosity 
halo subdwarfs are
{\sl not} found at colours comparable with those of their disk counterparts.
This fact has also been emphasised recently by Graff \& Freese (1996).
The field halo is a metal-poor population with an average abundance of
[Fe/H]$\sim$-1.5$\pm$0.3 (Kraft, 1989). Theoretical calculations by
VandenBerg et al (1983), D'Antona (1987), Burrows, Hubbard \& Lunine (1989)
and Baraffe et al (1995)
all indicate that stars of this abundance have both higher temperatures
and higher luminosities than solar-abundance stars of the same mass. 
Thus, Baraffe et al find that a 0.1 M$_\odot$ subdwarf with [Fe/H] = -1.5
has M$_{bol} = +12$ and T$_{eff} \sim 3300 K$, while a disk dwarf of the
same mass has M$_{bol} = +12.5$ and T$_{eff} \sim 2700 K$. These
differences are enhanced when converting to the observed (M$_V$, colour)
plane, since the lower line blanketing in the metal-poor stars leads to
a greater proportion of the total flux emitted in the optical region of
the spectrum, and hence both substantially bluer colours 
((V-I) $<$ 3.0) and higher optical luminosities.
\oneskip
These theoretical predictions are in good agreement with observations 
of both the lower main-sequence stars in globular clusters
(RF; Paresce et al, 1995) and of local parallax subdwarfs (Monet et al, 1992).
The faintest extreme subdwarf in the latter sample (LHS 1742a) has M$_V$ = 14.4
and (V-I) = 2.74, but M$_{bol} \sim 12$, comparable with the disk dwarf VB 10
( M$_V$=18.5, (V-I) = 4.33). Thus, counting the number of star at 
faint magnitudes with (V-I)$>$3.0 can probe only the metal-rich
([Fe/H] $< \sim -1$) tail of the halo abundance distribution and 
tells one essentially nothing about the bulk of the Galactic halo.
\oneskip
Second, it is important to consider the volume-sampling effects inherent
in any pencil-beam starcount analysis. As illustrated by RM93
(their figure 3), the convolution of a
monotonically-decreasing density law with the sampling 
volume of a pencil-beam survey leads to a `preferred' distance 
modulus for each stellar component, 
with that distance modulus (and the dispersion) dependent on the slope of
the density law along the particular Galactic line of sight. In the case
of a slightly-flattened (axial ratio $\sim 0.8$),
 r$^{-3.5}$ halo sampled at moderate to high latitude, the median distance
is 12.5 kiloparsecs (distance modulus 15.4 mag.). Fifty percent of
the halo stars lie between 8.2 and 21.0 kpc. (distance moduli 14.5 and 
16.6 mag. respectively) while 80 \% of the halo stars fall between 7.0 and 25.0
kpc. ((m-M) = 14.2 and 17.0 mag.). 
We emphasise that this is a {\sl general} property of starcounts - the 
spheroidal nature of the halo produces little variation in the
predicted number-magnitude distribution, even at low Galactic latitudes.
The result is that at magnitudes brighter than R$\sim$24, late-type 
subdwarfs (M$_V > 11$, M$_R > 10$) are outnumbered more than 
ten to one by more luminous halo stars lying closer to the main-sequence
turnoff.
\oneskip
Deep starcount surveys, however, reveal a significant number of 
moderately red ((V-I)$>$ 1.5 mag.) stars. What population do
these stars belong to? We have argued that late-type halo stars are rare,
thus the implication is that they are members of the disk population, and 
this hypothesis is supported by starcount predictions. There is still no
unanimity over the appropriate physical description of the old
disk/thick disk/intermediate population II (IP II) relationship, but provided
that one has an adequate representation of the overall density law 
perpendicular to the Plane (i.e. one that satisfies data from other surveys), 
that debate is unimportant for present purposes. 
In justification of this, we have computed starcounts towards the 
Galactic Pole for two models chosen to characterise two different interpretations
of the disk component. First, a model based on the RM93 discussions which
adopts exponential distributions for both old disk and IP II, with respective
scaleheights of 350 and 1500 pc. and a local normalisation IP II/old disk of
2 \%; second, and closer in construction to the Robin et al. (1996) Besancon
models, calculations based on adopting a sech$^2 {z \over z_0}$ density law
for the disk, with 
$z_0$ of 350 pc. for the old disk and 1500 pc. for IP II, with the local
normalisation set at 5.0 \%. (Note 
that the IP II constitutes close to 20 \% of the total mass of 
the disk in the latter model.) Exponential density laws have long been popular
in galaxy modelling, despite the lack of an obvious physical basis, 
while sech$^2$ distributions imply an isothermal population. 
Figure 7 plots the density distributions predicted by both models against the
observed density law derived from photometric parallax analysis of the
Gilmore \& Reid (1983) SGP starcounts, restricting the sample to 6.0 $< M_V <$ 7.0.
While there are substantial differences in the
relative proportions of the two components at any given height, in both
models the overall run of density gives a reasonable match to the observations. 
\oneskip
Since the density distribution of the disk populations falls off faster with
height above the Plane than does that of the halo, the disk 
stars lie at a smaller average distance. Looking towards the Galactic pole, 
the old disk stellar distribution peaks at (m-M)=9.0 and the IP II at
(m-M) = 12.5 in model A, while the respective distance moduli are 8.5 and 11
in model B. These distances scale with the inverse of the sine 
of the Galactic latitude for the disk populations. At high latitudes (b $> \sim 60^o$),
if one selects stars at a given apparent magnitude, say V=20,
model A predicts a sample that consists predominantly of
M$_V$=4.5 ($\pm \sim 1.0$) mag. halo subdwarfs, M$_V$=7.5 mag.
IP II dwarfs and M$_V$=11.0 mag. old disk dwarfs. Model B predicts 
M$_V \sim$9.0 mag. and M$_V \sim$11.5 mag. for the two disk populations. 
The net result is that at a given apparent magnitude, we expect the
average absolute magnitude of a disk star (either old disk or IP II) to
be significantly fainter than that of a halo subdwarf i.e. disk
dwarfs are redder than halo subdwarfs. This is the source of the bimodal colour
distribution in (B-V) at faint magnitudes originally detected by Kron (1980).
\oneskip
As regards the other parameters of our disk-population models, 
we have adopted an NGC 2420-like
([Fe/H]=-0.4) colour-magnitude diagram for the IP II (taking 47 Tuc as 
the template moves the colours to the blue by 0.05 magnitudes at most),
while the disk CMDs are based on nearby star data (Bessell, 1990). 
The halo population has been modelled using colour-magnitude relations
derived from observations of NGC 6752 (Thompson et al, in prep.), with the 
luminosity function taken from RM93 and an adopted local normalisation of
0.15 \% the local disk star density. 
Figure 8 shows the (R, (R-I)) colour-magnitude distributions predicted by
our models for two fields - (l=40, b=40) representing PSR 1640 and 
(l=110, b=75), close to the RF field. 
\oneskip
There are three important conclusions one can draw from these figures:
first, the colour-magnitude distributions graphically
demonstrate the systematic progression to fainter {\sl absolute} magnitudes 
(redder (R-I)) with
fainter {\sl apparent} magnitudes for each component. Second, the relative
invariance in the position of the halo colour-magnitude distribution is
clear, despite the substantial difference in galactic latitude. (There are
more stars in the lower latitude field since one is looking inward, past
the Galactic Centre.) Finally, it is clear
that even at R=24 magnitudes, the overwhelming majority of stars redder than
(R-I)=1.0 magnitude are contributed by the {\sl disk} populations. This
is the crucial circumstance that mitigates against the use of faint
starcounts to probe the halo luminosity function for late-type M subdwarfs.
In the following section we undertake a more detailed examination of the
predicted colour distribution at faint magnitudes, matching the models
against both the Keck data and the deeper but smaller field HST observations.
\oneskip
{\bf 4  A comparison between observations and model predictions}
\oneskip
{\sl 4.1 Ground-based observations}
\oneskip
Figure 9 matches the starcounts derived from the Keck images of field PSR 1640
against the predictions of the two models described in the previous section.
We have identified the contribution of the three main components in the 
latter models using separate symbols. Given the relatively small sample
size, either model provides an adequate fit to the data to at least R=24
magnitude. Both models clearly identify stars redder than (R-I)$\sim$1 as
from the disk - either an almost-equal mixture of old disk and IP II (model A)
or predominantly IP II (model B). This is important, since although RF
eliminated the extremely red stars from their sample ((V-I)$>$3.8), they
do not distinguish between halo subdwarfs and IP II stars. 
Their justification for this approach
(and, implicitly, for their including stars with heights above the Plane
as low as 1.25 kpc.) is based on arguments for a similarity between the
IP II and M71 colour-magnitude diagrams and, by inference, luminosity
functions. It is our proposition that this analysis is incorrect since the
starcount models show that one is combining halo and disk stars from
distinct, non-overlapping ranges in absolute magnitude. 
\oneskip
We can use our models to illustrate this directly for the high-latitude
RF field. Figure 10 plots the observed (V, (V-I)) distribution for the
stars identified by RF, together with the predicted colour-magnitude
diagram for a 0.0122 square-degree field at (l=109$^o$, b=73$^o$). We
predict 54 stars brighter than I=24.5, as compared with 49 observed.
The horizontal dotted line in the figure indicates the apparent 
magnitude limit of V=25; the vertical dotted line marks (V-I)=1.75. Our 
model predicts that {\sl all} of the $V < 25$ stars redward of the latter 
limit are from the disk (IP II in the case of model B). The solid points
identify stars which lie above 1.25 kpc (using the (M$_V$, V-I) relation
in RF) - all of the halo stars and 65 \% of the IP II. Thus, a luminosity 
function constructed from these data represents a composite based on blue halo
subdwarfs and red IP II stars, with the transition between the two at
(V-I)$\sim$ 1.7, or M$_V \sim$ 10.0 magnitudes (using the RF (V-I)
calibration). The reddest IP II stars are predicted to have (V-I)$\sim$2.8 mag.,
M$_V$(RF)$\sim$13.75, matching the faintest stars in the RF luminosity
function. Since the IP II has
a higher local density normalisation than the halo, one might expect a
luminosity (or mass) function derived from the composite starcounts to rise 
towards lower luminosities and smaller masses. Indeed, the
luminosity function derived by RF shows a sharp upturn in number density at
M$_V \sim +10$ - exactly the transition between the halo-dominated
and disk-dominated colour r\'egimes.
\oneskip
Any attempt to use starcounts to study low-luminosity stars in the halo
must take proper account of the contributions of {\sl both} old disk
and IP II at the appropriate colours. This demands both large areal
coverage (given the low surface-density of intrinsically-faint
subdwarfs at even R=25) and a more accurate knowledge of the 
density laws of the dominant disk population than we have at present.
\oneskip
{\sl 4.2 The Hubble Deep Field - starcounts}
\oneskip
\noindent
Ground-based starcounts are limited to magnitudes brighter than R=25 by the
coarse spatial filtering imposed by atmospheric seeing. The obvious method
of circumventing this problem is to rise above the Earth's atmosphere - a
condition which is met by the Hubble Space Telescope. The field of view
of the WFPC2 is extremely small (only 0.00157 sq. degrees), so the expected
number of stars in a typical `deep' single field is  small ($< 10$ to
R $\sim$ 26). Nonetheless, the recently-completed observations of the Hubble Deep 
Field (Williams {\sl et al}, 1996) permit the first test of starcount models 
at levels as faint as R=27 magnitude. 
\oneskip
The HDF observations consist of multi-orbit exposures in 
four filters (F300, F450, F606, F814) of a single low-reddening
WFPC2 field at l=126$^o$ and b=55$^o$. The latter two filters are
the most interesting for starcount work and, as in the study by 
Flynn, Gould \& Bahcall (1996 - FGB), we have restricted our analysis to
these two bands, which have effective limiting magnitudes of
V $\sim$ 30 and I $\sim$ 30 on the Johnson/Cousins system. 
We have used the SExtractor package to determine magnitudes and 
morphological parameters for all of the sources in the V- and I-band frames,
and all of the objects classed as stellar were also checked by eye.
The magnitude zeropoints are as specified by STScI, adopting the
aperture corrections and colour-terms (particularly significant
in F606W) given by Holtzman et al (1995). Saturation limits our sample
to stars fainter than I $\sim$ 18.5 magnitude. 
\oneskip
Our star/galaxy classification is based on a single technique - plotting
the peak brightness against the total magnitude measured on the F814W
frames. The position of the stellar sequence is defined by taking a
bright (unsaturated) star as a PSF template and scaling the intensities
(adding the appropriate noise) to match stars at fainter magnitudes. 
The results are plotted in figure 11, where the simulated stellar
sequence is plotted as open circles; objects classed as stars are plotted
as solid triangles; probable stars are plotted as solid squares; and
galaxies are plotted as crosses. 
The benefit of 0.1 arcsecond `seeing'
is obvious - the stellar sequence is clearly distinct from the galaxy 
population to at least I=26 magnitudes, and we would predict
relatively little contamination even at I=27.
This represents an effective gain of at least 2 magnitudes
over ground-based R-band observations.
\oneskip
As expected, however, the total number of stars is very small. Out of a total
of $\sim 700$ objects, we find
only 16 definite and 4 probable stars in the range 20 $< I <$ 27.
Figure 12 plots the (I, (V-I)) colour-magnitude diagram for these objects.
This can be compared directly with the colour-magnitude diagram derived
by Flynn {\sl et al} (1995) (their figure 2), results obtained using entirely
independent, and more sophisticated, methods. We find one 
additional faint red star 
(I=25.7, V-I=2.3 - lying within the FGB selection zone), but otherwise 
the two diagrams are identical for 20 $< I <$ 26 magnitudes - a 
testament to the robustness of the morphological classification techniques.
\oneskip
Figures 13a and 13b show the (I, (V-I)) colour magnitude diagrams predicted
by our models A and B. The latter were computed for a
solid angle of 0.0157 sq. deg. - ten times the HDF solid angle - to
better display the relative contributions of the different stellar
populations. Note how the contribution from the Galactic disk (either
old disk or IP II) peters out at I$\sim$ 26 magnitudes in both
models as we reach the bottom of the (metal-rich) hydrogen-burning 
main-sequence. Only at fainter magnitudes is the halo
the dominant stellar population at {\sl all} colours.
However, even the HDF data are incapable
of discriminating between a continuously increasing halo luminosity function 
(RF) and the DLHG function. At I$\sim$ 26, the typical
halo subdwarf has (V-I)$\sim 2$ and M$_V \sim 11 \pm 1$, while the two luminosity
functions diverge only at M$_V > +12$. In general, the total 
starcounts predicted by
both models are in reasonable agreement with the observations - 
for 20 $<$ I $<$ 26, model A predicts 10 stars (1.5 disk, 2.5 IP II, 
6 halo) while model B predicts 13 stars (0 disk, 7 IP II, 6 halo).
This compares with the 13 to 15 stars observed in the same magnitude
range. 
\oneskip
Similarly, the observed and predicted colour distributions are
in reasonable agreement, with the redder stars identified as disk or
IP II and the intermediate-colours objects as halo subdwarfs. 
There is a marginal deficit of faint, red stars - both models
predict $\sim 4$ stars with (V-I) $>$ 1.0 magnitudes and 25 $<$ I $<$ 27,
while only two stars are observed (one of which may be a misclassified
galaxy). Clearly, further observations covering more fields are required 
to establish whether this is anything more than small number statistics.

\oneskip
{\sl 4.3 The Hubble Deep Field - constraints on baryonic dark matter}
\oneskip
Flynn et al  have used the absence of faint, red stars in the HDF starcount data to 
set limits on the baryonic (stellar) contribution to the dark-matter halo (DMH).
They derive upper limits of 6 \% for objects with M$_I < 15$ and 1 \% for
M$_I < 14$. However, while these luminosities are appropriate for very 
low-mass ($\sim 0.1 M_\odot$) solar abundance, disk dwarfs, there are
reasonable grounds for assuming that they are not appropriate for the
hypothetical DMH dwarfs. Given that the DMH has an extended (spheroidal)
distribution, it is probable that it formed before any substantial dissipational 
collapse occurred in the proto-galaxy. Hence, it also seems likely
that any stellar constituents would have metal abundances 
at least comparable with the conventional Galactic halo - a point also
made by Graff \& Freese (1996).
As described in section 3, the lower blanketing at lower
abundances leads to brighter optical luminosities than amongst low-mass disk dwarfs.
Thus, a better template for the hypothetical DMH dwarfs is given either by the 
zero-metallicity models computed by Saumon et al (1994), who find M$_V$=12.8 and
(V-I)=1.6 for a 0.092 M$_\odot$ object (the minimum mass for hydrogen burning
at this abundance), or by the extreme subdwarfs (M$_V$=14.5, (V-I)=2.7)
included in the Monet et al. parallax stars. (For reference, a 0.1 M$_\odot$ 
solar-abundance disk dwarf has M$_V \sim$ +19, (V-I)$\sim$ 4.5 magnitudes.)
Under either of these assumptions we can set significantly 
more stringent limits on the baryonic contribution to the DMH.
\oneskip
First, surveys of the local stars are complete to distances of $\sim$14
parsecs for stars of M$_V = 13 \pm 0.5$ and $\delta > -30^o$ (Reid, Hawley \& Gizis, 1995).
If we ascribe all of the local DMH density ($\sim 0.009 M_\odot$ pc$^{-3}$ -
FGB) to stars at the H-burning limit, we have a local density of 0.1 DMH stars pc$^{-3}$,
or $\sim 860$ stars within the 14 parsec distance limit appropriate for
zero-metallicity stars. None have been found,
so we can infer that no more than $\sim 0.13 \%$ of the mass of the DMH rests
in the form of zero-metallicity M-dwarfs.
On the other hand, we can take LHS 1742a (M$_V$=14.4, (V-I)=2.74 magnitudes)
as our DMH dwarf template. Nearby stars surveys are complete only within a
distance of 7 parsecs for stars of this absolute magnitude.  
Again, there are no plausible candidates currently
known, limiting the contribution of these stars to $\sim$1 \% of the
local dark-matter density. 
\oneskip
Second, if we assume that
the DMH has an r$^{-2}$ distribution, we can use the HDF starcounts to tighten
these limits. With a local density of 0.1 DMH stars pc$^{-3}$, we would 
expect $\sim$7000 
DMH dwarfs of 0.092 M$_\odot$ brighter than I=26.25 in the HDF sample. We
observe three stars with the appropriate colours ( (V-I) = 1.6 $\pm 0.2$ mag.),
implying that no more than 0.04 \% of the dark matter
halo can be found as zero-metallicity, main-sequence stars near the
hydrogen-burning limit. Alternatively, if we
take LHS 1742a as a DMH dwarf template, then we predict 2600 DMH dwarfs in the
HDF field. Again, there are only three candidates within the appropriate
colour range, limiting the contribution of
low-mass, metal-poor subdwarfs to no more than 0.12 \% of the total DMH mass.
These results are similar to those derived by Graff \& Freese (1996) based
on re-analysis of the Bahcall et al (1994) HST data and appear to rule 
out a significant contribution to the DMH by hydrogen-burning,
main-sequence stars of any abundance.
\oneskip
The only remaining stellar option for the dark-matter halo is white
dwarf stars. Our HDF colour-magnitude diagram includes a cluster of
three blue objects at I $\sim$ 26, (V-I) $\sim$ 0.5 magnitudes. 
These objects are also identified as being stellar in morphology by FGB,
although they derive slightly bluer colours. 
White dwarf stars, mainly from the disk, are the only objects predicted at
these colours and luminosities - but at significantly lower surface
densities. Based on our models, one expects 0.2 disk white dwarfs and 0.01 halo 
white dwarfs per
WFPC field. Thus, the most conservative conclusion is that these faint, blue,
compact objects are either low-luminosity active galactic nuclei
or barely-resolved cores in faint, star-forming galaxies.
However, the most recent analysis of the (relatively few) MACHO detections 
gives a most probable mass for the lensing objects of $\sim 0.5 M_\odot$
(Alcock et al, in prep.) - consistent with white dwarfs. Are the
I$\sim$26 mag. blue objects consistent with expectations if a 
significant fraction of the mass of the dark-matter halo is in the
form of white dwarfs?
\oneskip
If we interpret the 26th magnitude objects as white dwarfs, then
the observed colours imply temperatures of $\sim$ 10,000 K and 
relatively short cooling times of $< 10^9$ years. Our observations
rule out any substantial contribution to starcounts from lower-mass,
main-sequence counterparts of these white dwarfs, so we have to
postulate both a sharp break in the initial mass function and
fine tuning such that stars above the break evolved only recently 
onto the white dwarf sequence. If these white dwarfs have an average
mass of 0.6 M$_\odot$, then a local number density of 0.015 M$_\odot$
ps$^{-3}$ is sufficient to match the inferred local DMH mass density.
If we further assume that the observed objects
mark the upper end of a luminosity function similar in shape to the local
disk white dwarf function (Liebert, Dahn \& Monet, 1988), 
then we predict 32 stars with (V-I) $<$ 1.2 mag. and 
23 $<$I $<$ 26 in the HDF field and a further 60 with 26 $<$ I $<$ 27
magnitudes. Figure 12 includes only three stars in the relevant colour
range with I$<$26 mag., implying a contribution of no more than
10 \% to the dark-matter halo mass in this model - and probably 
substantially less, since all three observed stars are more likely to be 
main-sequence subdwarfs in the conventional halo. 
\oneskip
There remains at least one model for an all-baryonic dark-matter halo which
is consistent with the low number density of stars at
brighter magnitudes in the HDF. If we
construct the dark-matter halo solely from old, low-luminosity (M$_V = 16$) 
white dwarfs, the we predict only one DMH white dwarf
with I$<$26 and only a further nine objects with 26 $< I <$ 27. All would
have (V-I) colours of 1.2 to 1.5 magnitudes. However, the predicted
counts rise steeply towards fainter magnitudes, with a maximum surface
density of $\sim 450$ stars per WFPC field at I$\sim 32 \pm 0.5$. 
We would expect 70 objects in the HDF at I=28.5 $\pm 0.5$ mag. (approximately
10 \% of the galaxy number counts at that magnitude) and $\sim$300
at I=30.5 $\pm 0.5$ magnitudes. The major drawback, of course, is that 
placing even a large fraction of the mass of the dark-matter halo into 
these old, low-luminosity white dwarfs
requires extreme fine-tuning in the initial mass function to minimise
both the number of high-mass (M $>$ 4 M$_\odot$) stars (and over-enrichment of the 
stellar halo from planetary nebulae and supernovae ejecta) and the number of 
long-lived, low-mass ($<$ 0.8 M$_\odot$) main-sequence stars. 
\oneskip
In summary, while these results demonstrate the substantial advantage in
image classification afforded by the increased resolution of HST data, they
also highlight the major limitation - the small field of view. One can
use data from a single field to constrain the possible constituents of 
an r$^{-2}$ dark-matter halo, as FGB previously demonstrated, but the sparser 
luminous stellar populations of the disk and halo require observations
covering at least a factor of ten more in solid angle. 
\oneskip
{\bf 4. Summary and conclusions}
\oneskip
\noindent
We have rediscussed the technique of using faint starcounts as a probe of the
halo subdwarf luminosity function. We have shown that even good-seeing 
ground-based observations are limited to $R \le 25.5$ by star/galaxy separation
problems - problems introduced by the intrinsically small size of most
galaxies at these faint magnitudes. HST data can extend accurate classification
at least two magnitudes fainter, but each WFPC field provides only a handful of stars.
Given the small area accessible to adaptive
optics correction and the low surface density of (all) faint stars, 
it is unlikely that such techniques will be of much practical benefit in the
near future, and wide-field space-based observations are probably the most
effective means of probing to such faint magnitudes.
\oneskip
At brighter magnitudes, star/galaxy separation techniques can be applied
ground-based observations taken under good seeing conditions, although 
with variable success rates. We have
examined four different methods and show that, depending on the method 
adopted and how conservative the final classification is, the number of 
objects classified as stars can vary by up to $\pm10$ \% in the magnitude 
interval $23 < R < 24$. 
\oneskip
Turning to analysis of the post-classification stellar sample, we have used
starcount models to show, first, that fewer than 60 \% of the stars with
$20 < R < 24$ are from the Galactic halo and, second, that the vast majority
of these subdwarfs are brighter than $M_V = +8$. The red stars 
((V-I) $> \sim$2) at these magnitudes are contributed by the disk, with the
relative contributions of the old disk and IP II dependent on the 
prescription adopted for the deconvolution of the disk. The lowest-luminosity halo
subdwarfs, the majority of which have (V-I) $<$ 3.0 magnitudes,
lie at substantially fainter magnitudes ($I > 28$). We have
also compared our model predictions with starcounts from the Hubble Deep Field
and find general agreement - although the observed stellar sample is extremely
small. The HDF data do, however, allow one to set constraints on several
possible baryonic contributors to the dark-matter halo. 
\oneskip
Deep starcounts are most directly applied to studies of the luminous
stellar populations in the Galaxy.
Based on the results presented in this paper, 
we conclude that the steep luminosity/mass function deduced by 
Richer \& Fahlman (1992) for the field stars in the Galactic halo is a 
result of combining observations of intrinsically bright halo stars with 
intrinsically faint (and more numerous) IP II stars. We would argue that
there is thus no evidence that the mass function of the halo population
is continuously increasing towards the hydrogen-burning limit. In general,
deep ground-based starcounts are not an effective means of probing the
faint end of the luminosity function of the stellar Galactic halo.
\oneskip
Acknowledgements: we would like to thank J. G. Cohen for taking the
Keck observations and S.R. Kulkarni for making these data available to 
us. LY also acknowledges useful discussions with Warrick Couch and
Ren\'e Mendez. We also acknowledge useful comments by the referee,
Harvey Richer, and thank him for making available the VRI photometry
of stars in the RF CFHT field. The Keck telescope
project was made possible by a generous grant from the W.M. Keck
Foundation. This research was supported partially by NSF grant AST-9412463.
\vfill\break
\noindent
\oneskip
{\bf References }
\oneskip
\hyphenpenalty=10000
\def\ref#1\par{\parshape=2.0in 15.2 truecm 1.0truecm 14.2 truecm {#1} \par}
\parskip=0pt
\parindent=0pt

\ref
Anderson, J., King, I.R. 1996, {A.S.P. Conference Series vol. 92}, {257},
(ed. H. Morrison \& A. Sarajedini)

\ref
Bahcall, J., Flynn, C., Gould, A., Kirhakos, S. 1994, {\sl ApJ}, {\bf 435},
L51

\ref
Baraffe, I., Chabrier, G., Allard, F., Hauschildt, P. 1995, {\sl ApJ},
{\bf 446}, L35

\ref
Bertin, E. 1995, {\sl SExtractor manual}, {IAP}, Paris

\ref
Bessell, M.S. 1990, {\sl AAS}, {\bf 83}, 357

\ref
Burrows, A., Hubbard, W.B., Lunine, J.I. 1989, {\sl ApJ}, {\bf 345}, 939

\ref
Dahn, C.C., Liebert, J., Harris, H.C., Guetter, H.H., 1994, in
{\sl The Bottom of the Main Sequence - and beyond}, {Springer-Verlag,
Berlin}, ed. C. G. Tinney, p. 239

\ref
D'Antona, F., 1987, {\sl ApJ}, {\bf 320}, 653

\ref
Drukier, G.A., Fahlman, G.G., Richer, H.B., Searle, L., Thompson, I.
1993, {\sl AJ}, {\bf 106}, 2335

\ref
Elson, R.A.W., Gilmore, G.F., Santiago, B.X., Casertano, S. 1995, {\sl AJ},
{\bf 110}, 682

\ref
Flynn, C., Gould, A., Bahcall, J., 1996, {\sl preprint}, {}

\ref
Gilmore, G.F., Reid, I.N. 1983, {\sl MNRAS}, {\bf 202}, 1025

\ref
Graff, D.S., Freese, K. 1996, {\sl ApJ Lett.}, {in press}

\ref
Holtzman, J.A., Burrows, C.J., Casertano, S., Hester, J.J., Trauger, J.T.,
Watson, A.M., Worthey, G. 1995, {\sl PASP}, {\bf 107}, 1065

\ref
Jones, L.R., Fong, R., Shanks, T., Ellis, R.S., Peterson, B.A. 1991,
{\sl MNRAS}, {\bf 249}, 481

\ref
Kirkpatrick, J.D., McGraw, J.T., Hess, T.R., Liebert, J., McCarthy, D.W.
1994, {\sl ApJS}, {\bf 94}, 749

\ref
Kraft, R.P. 1989, {\sl PASP}, {\bf 101}, 1113

\ref
Kron, R.G., 1980, {\sl ApJS}, {\bf 43}, 305

\ref
Kroupa, P., 1995, {\sl ApJ}, {\bf 453}, 358

\ref
Landolt, A. U. 1992, {\sl AJ} {\bf 104}, 340

\ref
Liebert, J., Dahn, C.C., Monet, D.G. 1988, {\sl ApJ} {\bf 332}, 891

\ref
Luyten, W.J., 1979, {\sl LHS Catalogue}, {Univ. of
Minnesota}, Minnesota

\ref
Metcalfe, N., Shanks, T., Fong, R., Roche, N. 1995, {\sl MNRAS}, 
{\bf 273}, 257

\ref
Monet, D.G., Dahn, C.C., Vrba, F.J., Harris, H.C., Pier, J.R., 
Luginbuhl, C.B., Ables, H.D. 1992, {\sl AJ}, {\bf 103}, 638

\ref
O'Connell, D.J.K. 1958, editor, {\sl Specola Vaticana}, {\bf 5}

\ref
Oke, J. B., Cohen, J. G., Carr, M., Cromer, J., Dingizian, A., 
Harris, F. H., Labreque, S., Lucinio, R., Schaal, W., Epps, H. 
\&\ Miller, J. 1995, {\sl PASP}, {\bf 107}, 375

\ref
Paresce, F., Demarchi, G., Romaniello, M. 1995, {\sl ApJ}, 
{\bf 440}, 216

\ref
Reid, I.N., Majewski, S.R. 1993, {\sl ApJ}, {\bf 409}, 635 (Paper I)

\ref
Reid, I.N., Hawley, S.L., Gizis, J.E. 1995, {\sl AJ}, {\bf 110}, 1838

\ref
Richer, H., Fahlman, G., Buonnano, R., Fusi Pecci, F., Searle, L., 
Thompson, I. 1991, {\sl ApJ}, {\bf 381}, 147

\ref
Richer, H.B., Fahlman, G.G. 1992, {\sl Nature}, {\bf 358}, 383

\ref
Robin, A.C., Haywood, M., Cr\'ez\'e, M., Ojha, D.K., Bienaym\'e, O., 
1996, {\sl AA}, {\bf 305}, 125

\ref
Santiago, B.X., Gilmore, G., Elson, R.A.W. 1996, {\sl MNRAS}, {in press}

\ref
Saumon, D., Bergeron, P., Lunine, J.I., Hubbard, W.B., Burrows, A. 1994,
{\sl ApJ}, {\bf 424}, 333

\ref
Smail, I. R., Hogg, D. W., Yan, L. \&\ Cohen, J. G. 1995, {\sl ApJL}, 
{\bf 449}, L105

\ref
Stark, A.A., Gammie, C.F., Wilson, R.W., Bally, J., Linke, R.A.,
Heiles, C., Hurwitz, M. 1992, {\sl ApJS}, {\bf 79}, 77

\ref
Stetson, P.B. 1987, {\sl PASP}, {\bf 99}, 613

\ref
Tinney, C.G. 1993, {\sl ApJ}, {\bf 414}, 254

\ref
VandenBerg, D. A., Hartwick, F.D.A., Dawson, P., Alexander, D.R. 1983,
{\sl ApJ}, {\bf 266}, 747

\ref
Williams, R. et al 1996, {Science with the Hubble Space Telescope II}, 
{P. Benvenuti, F.D. Macchetto \& E.J. Schreier, eds}, in press
(Baltimore, STScI)

\vfill\break
\centerline {\bf Figure Captions}
\oneskip
$Figure~1.$ A comparison between galaxy counts and star counts at high
galactic latitude. The solid points are the galaxy counts, taken from
Metcalfe et al (1995) and S95. The lines represent the predicted
star counts at [l=90$^o$, b=45$^o$] (dotted line), [l=90$^o$, b=60$^o$] 
(dashed line) and the Galactic Pole (solid line). These have been calculated
using model B (described in section 3 of the text).
\oneskip
$Figure~2.$ a: the (V-R)/(R-I) distribution of nearby disk stars (Bessell,
1991) and halo subdwarfs. b: The VRI distribution of galaxies in the
PSR 1640 field. The field-star sequence is also plotted, and the solid lines
mark offsets from that sequence of $\pm$ 0.14 magnitudes in each colour.
\oneskip
$Figure~3.$ The upper two panels show the distribution of image ellipticity
as a function of apparent magnitude in the two Keck fields while the
lower panel shows the predicted stellar distribution, based on 200 
artificially-generated stars. The horizontal line marks the star/galaxy
classification threshold adopted.
\oneskip
$Figure~4.$ The distribution of the DAOPHOT $\chi$ parameter as a function
of magnitude. Again, the upper two panels show the observed distributions
in the two fields and the lower panel shows the distribution predicted by
the 200 artificial stars. The heavy solid lines in figures 3a and 3b outline 
the stellar locus, and the horizontal line marks $\chi=3$. All objects
lying above the latter line were classed as galaxies. 
\oneskip
$Figure~5a.$ The observed distribution in the PSR 1640 field 
of the Jones et al (1991) 
peak-brightness parameter used by RF as their only star/galaxy criterion. 
The lower figure shows the distribution predicted by the artificial
stars.
\oneskip
$Figure~5b.$ The observed distribution in PSR 2229 of the Jones et al (1991) 
peak-brightness parameter used by RF. 
\oneskip
$Figure~6a.$ The ratio between the peak brightness (I$_P$) and the total
flux within a 0$"$.6 diameter aperture for sources in the PSR 1640 field.
As with the
Jones et al index, this measures the degree of central concentration of the
light distribution. Again, the lines plotted on the upper diagram represent
the 3-$\sigma$ limits on the stellar distribution defined by the artificial
stars plotted in the lower panel.
\oneskip
$Figure~6b.$ As figure 5a but for the PSR 2229 field.
\oneskip
$Figure~7.$ A comparison between the observed density distribution of K dwarfs
(6.0 $\le M_V \le$ 7.0) in the SGP (RM93) and the predicted distributions
of the two stellar models described in the text. The solid points show the
observed starcounts; the solid line marks the total starcounts predicted by
either model; the dotted line represents the contribution from the old disk;
the short-dashed line the contribution from the IP II; and the longer-dashed 
line the predicted number density of halo stars.
\oneskip
$Figure~8a.$ The (R, (R-I)) colour-magnitude distribution predicted by models A
and B for a 0.05 square degree field at (l=40$^o$, b=40$^o$). The halo
population has identical parameters in both models.
\oneskip
$Figure~8b.$ The (R, (R-I)) colour-magnitude distribution predicted by models
A and  B for a 0.05 sq. deg. high galactic latitude (l=110$^o$, b=75$^o$) 
field. Note that the colour-magnitude distribution of the halo stars is 
very similar to that in the lower latitude field. In both fields the 
lowest luminosity halo stars make a significant contribution to the 
starcounts only at magnitudes fainter than R$\sim$27.
\oneskip
$Figure~9a.$ A comparison between the observed (R, (R-I)) colour-magnitude diagram 
in the Keck PSR 1640 field and the predictions of model A. Halo subdwarfs in the
latter model are identified as solid points; IP II stars as open triangles and 
disk dwarfs as crosses. The same symbols are used in identifying the contributions
of the individual populations to the predicted histogram distributions at each
magnitude.
\oneskip
$Figure~9b.$ As figure 9a, but showing the predictions for model B. These are in
slightly better accord with the observations.
\oneskip
$Figure~10.$ The observed (V, (V-I)) colour-magnitude for stars in the RF CFHT field
and the distribution predicted by model
B for the 0.0122 square degree RF high latitude field. The stellar populations
are coded as in previous figures, with the solid triangles identifying IP II
stars lying more than 1.25 kpc. above the Plane (using the RF photometric
parallax calibration). The horizontal dotted line marks the
magnitude limit of V=25 while the vertical line marks a (V-I) colour of
1.7, corresponding to M$_V$=+10. All of the stars in the upper right 
quadrant are drawn from the IP II.
\oneskip
$Figure~11.$ Star-galaxy discrimination in the HDF field, plotting peak flux against
magnitude for all objects identified on the F814 I-band data. We have used a 
bright unsaturated star as a PSF template and the open circles, derived by
scaling this template to fainter magnitudes, define the stellar sequence. Solid
triangles are objects classed as definitely stellar; solid squares are possible
stars; crosses are galaxies. 
\oneskip
$Figure~12.$ The (I, (V-I)) colour-magnitude diagram described by objects classed
as stars in the Hubble Deep Field. The crosses identify objects where the
morphological classification is uncertain.
\oneskip
$Figure~13.$ The (I, (V-I)) colour-magnitude diagrams predicted by models A and
B for a solid angle ten times that of the HDF. The stellar populations are
coded as in figure 8.
\nopagenumbers
\vfill\eject
\centerline{{\bf Table 1.} Summary of Observations}
%\vfill\eject
\def\pz{\phantom{0}}
\magnification=\magstep1
\hoffset=0.0 true cm
\voffset=0.0 true cm
\vsize=22.5 true cm
\hsize=16.0  true cm
%\pageno = 11

%\special{landscape}
\baselineskip=12pt
\parskip=0pt
\parindent=12pt
%\raggedbottom

\null
\settabs
\+Fieldxxxxxxxx&
alpxxxxxxxxxxx&
delxxxxxxxxxxx&
Filxxxxx&
Expxxxxx&
Numbxx&
FWHMxxx&
.\cr

\hrule
\hrule
\smallskip
\hrule
\hrule
\medskip

\+ Field \hfill & \pz\pz\pz\pz $\alpha$ \hfill & \pz \pz\pz\pz $\delta$ \hfill & Filter \hfill & $\rm T_{tot}$ \hfill 
& N \hfill & FWHM \hfill & \cr 
\+ \hfill & \pz\pz J2000 \hfill & \pz\pz J2000 \hfill & \hfill & \pz sec \hfill & \hfill & \pz\pz $''$ \hfill & \cr
\medskip
\hrule
\hrule
\bigskip

\+ 1640+22 \hfill & 16$^h 40^m 18''.9$ \hfill & $+22^\circ24'19''.0$ \hfill & \pz V \hfill & 1500 \hfill & 1 \hfill & 0.78$''$ & \cr
\+ \hfill & $l=41.1^\circ$ \hfill & $b=38.3^\circ$ \hfill & \pz R \hfill & 2400 \hfill & 2 \hfill & $0.55''$ \hfill & \cr
\+ \hfill & \hfill & \hfill & \pz\ I \hfill & 2000 \hfill & 4 \hfill & $0.53''$ \hfill & \cr 
\smallskip
\hrule
\medskip
\+ 2229+26 \hfill & 22$^h29^m50''.9$ \hfill & $+26^\circ43'52''.8$ \hfill & \pz V \hfill & \pz900 \hfill & 1 \hfill & $0.87''$ \hfill & \cr
\+ \hfill & $l=87.7^\circ$ \hfill & $b=-26.3^\circ$ \hfill & \pz R \hfill & 2100 \hfill & 2 \hfill & $0.58''$ \hfill & \cr
\+ \hfill & \hfill & \hfill & \pz\ I \hfill & 1000 \hfill & 2 \hfill & $0.58''$ \hfill & \cr

\bigskip
\hrule
\hrule
\smallskip
\hrule
\hrule
\bigskip

\centerline {Notes to Table 1}
\noindent{The full-width half-maximum listed is the seeing from the final averaged image.
T$_{tot}$ is the total exposure time for each colour.}

\vfill\eject

\centerline{{\bf Table 2a.} Summary of Star Counts in Field PSR 1640}
%\vfill\eject
\settabs
\+methodXXXXXXXxxx&
R2xxxxxxxxx&
R3xxxxxxxxx&
R4xxxxxxxxx&
R5xxxxxxxxx&
.\cr

\medskip
\hrule
\hrule
\smallskip
\hrule
\hrule
\medskip

\+ Star \hfill & R=21$-$22 \hfill & R=22$-$23 \hfill & R=23$-$24 \hfill 
& R=24$-$25 \hfill & \cr 
\vskip3pt
\+ Selection \hfill & \ \ $\rm N_{star}$ \hfill & $\rm N_{star}$ \hfill & $\rm N_{star}$\hfill & $\rm N_{star}$ \hfill & \cr
\medskip
\hrule
\hrule
\bigskip

\+ $\epsilon$ \hfill & 42 \hfill & 79 \hfill & 228 \hfill & 905 \hfill & \cr
\+ $\chi$ \hfill & 58 \hfill & 142 \hfill & 290 \hfill & 459 \hfill & \cr
\+ $\epsilon$+$\chi$+$I_{peak}$ (R)\hfill & 34 \hfill & 41 \hfill & 54 \hfill & 107 \hfill & \cr 
\+ $\epsilon$+$\chi$+$F_{0.6}$ (R)\hfill & 36 \hfill & 35 \hfill & 51 \hfill & 96 \hfill & \cr
\+ $\epsilon$+$\chi$+$I_{peak}$+$F_{0.6}$(R) \hfill & 33 \hfill& 36 \hfill & 44 \hfill& 76 \hfill& \cr
\+ $\epsilon$+$\chi$+$I_{peak}$ (RI)\hfill & 33 \hfill & 32 \hfill & 51 \hfill & 95 \hfill & \cr 
\+ $\epsilon$+$\chi$+$I_{peak}$+$F_{0.6}$(RI) \hfill & 30 \hfill& 33 \hfill & 44 \hfill& 75 \hfill& \cr
\+ $\epsilon$+$\chi$+$I_{peak}$(VRI) \hfill & 30 \hfill& 32 \hfill &45 \hfill& 28 \hfill &\cr
\+ $\epsilon$+$\chi$+$I_{peak}$+$F_{0.6}$(VRI) \hfill & 27 \hfil& 33 \hfill & 38 \hfill& 23 \hfill& \cr
\vskip10pt
\hrule
\vskip10pt
\+ $\rm N_{galax}$ \hfill & 99 \hfill & 208 \hfill & 435 \hfill & 911 \hfill & \cr
\bigskip
\hrule
\hrule
\smallskip
\hrule
\hrule
\bigskip

\settabs
\+methodXXXXXXXxxx&
R2xxxxxxxxx&
R3xxxxxxxxx&
R4xxxxxxxxx&
R5xxxxxxxxx&
.\cr
\vfill\eject
\centerline{{\bf Table 2b.} Summary of Star Counts in Field PSR 2229}
%\vfill\eject
\medskip
\hrule
\hrule
\smallskip
\hrule
\hrule
\medskip

\+ Star \hfill & R=21$-$22 \hfill & R=22$-$23 \hfill & R=23$-$24 \hfill 
& R=24$-$25 \hfill & \cr 
\vskip3pt
\+ Selection \hfill & \ \ $\rm N_{star}$ \hfill & $\rm N_{star}$ \hfill & $\rm N_{star}$\hfill & $\rm N_{star}$ \hfill & \cr
\medskip
\hrule
\hrule
\bigskip

\+ $\epsilon$ \hfill & 80 \hfill & 141 \hfill & 228 \hfill & 461 \hfill & \cr
\+ $\chi$ \hfill & 69 \hfill & 147 \hfill & 332 \hfill & 1057 \hfill & \cr
\+$\epsilon$+$\chi$+$I_{peak}$ (R)\hfill & 45 \hfill & 75 \hfill & 99 \hfill & 203 \hfill & \cr 
\+$\epsilon$+$\chi$+$I_{peak}$+$F_{0.6}$ (R)& 42 & 64 & 46 & 82 & \cr
\+$\epsilon$+$\chi$+$I_{peak}$(RI) & 16 & 19 & 16 & 31 & \cr
\+$\epsilon$+$\chi$+$F_{0.6}$(RI) \hfill & 16 \hfill & 24 \hfill & 36 \hfill & 90 \hfill & \cr
\bigskip
\hrule
\hrule
\smallskip
\hrule
\hrule
\bigskip

\centerline {Notes to Table 2}
\noindent{As described in the text, we have used four techniques
to discriminate stars and galaxies. Column 1 identifies the criterion or
criteria used to derive the starcounts $-$ thus `$\epsilon$+$\chi$+$I_{peak}$' means that to
be classified as stars, objects must satisfy the ellipticity, $\chi$ 
and Jones et al (1991) criteria (this matches the classification 
method used by RF). We have listed the starcounts if we base our analysis
only on the R-band frames (identified as (R) in the Table); if we insist
on detection on the I-band
frames ( (RI) ); or detection in all three passbands ( (VRI) ).
Note that we require only detection in the other passbands - insisting on
classification as stellar would reduce the sample size still further.
Finally, we list the galaxy counts derived by S95 for field PSR 1640.
These last data have been corrected for incompleteness (see S95
for full details).}
\vfill\eject
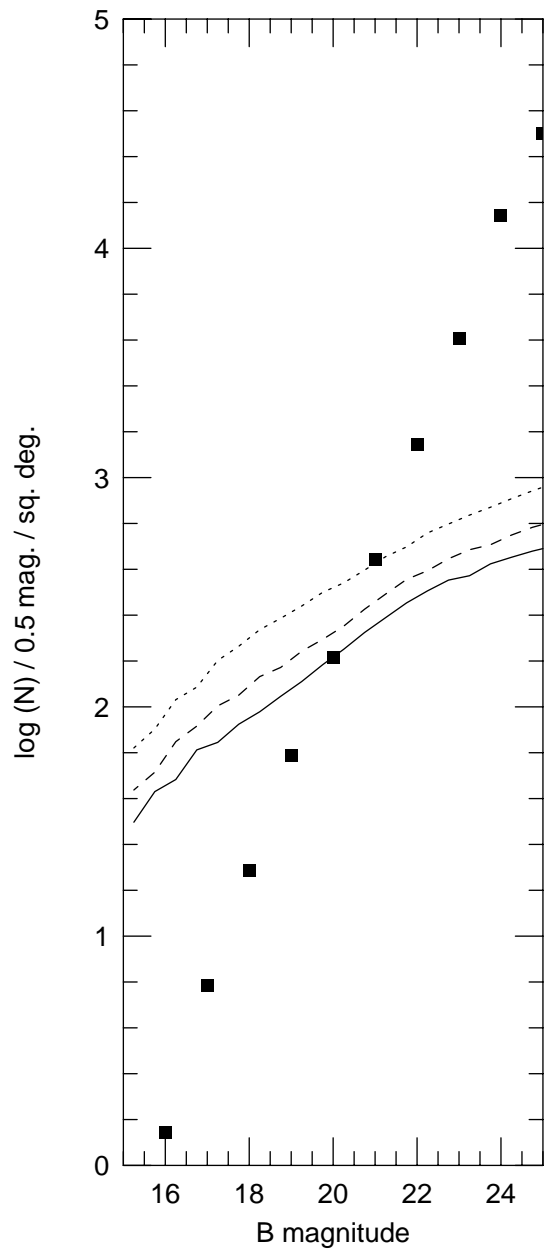
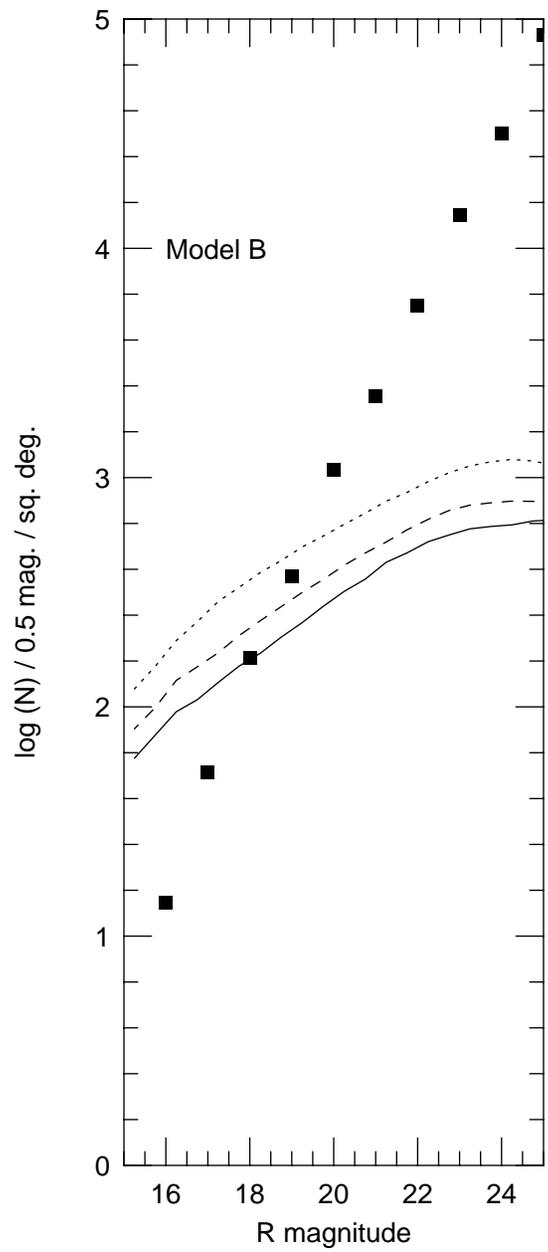
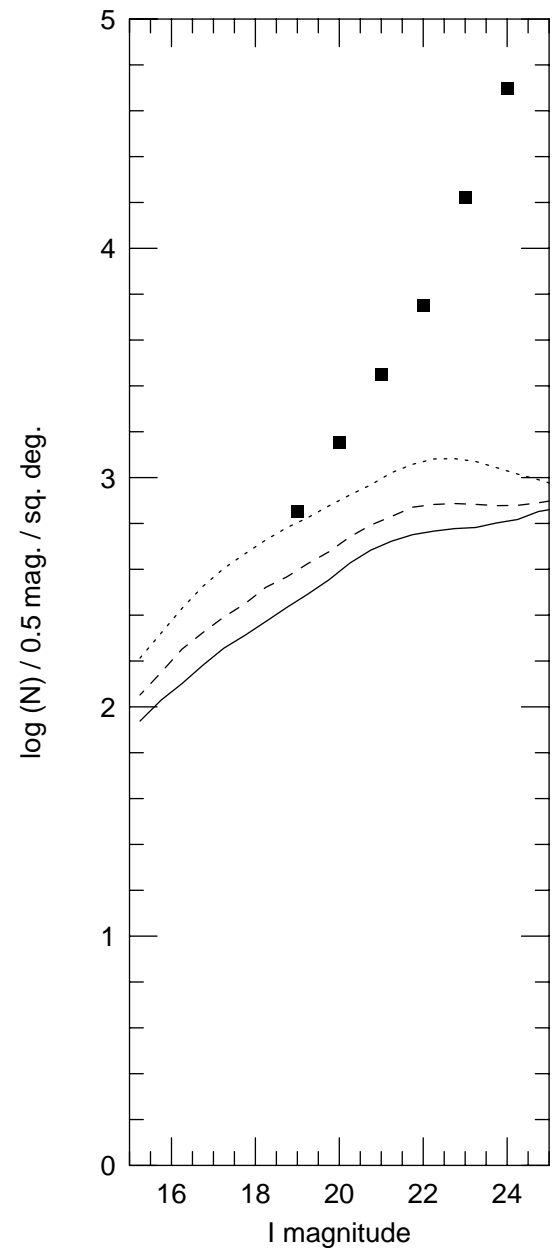
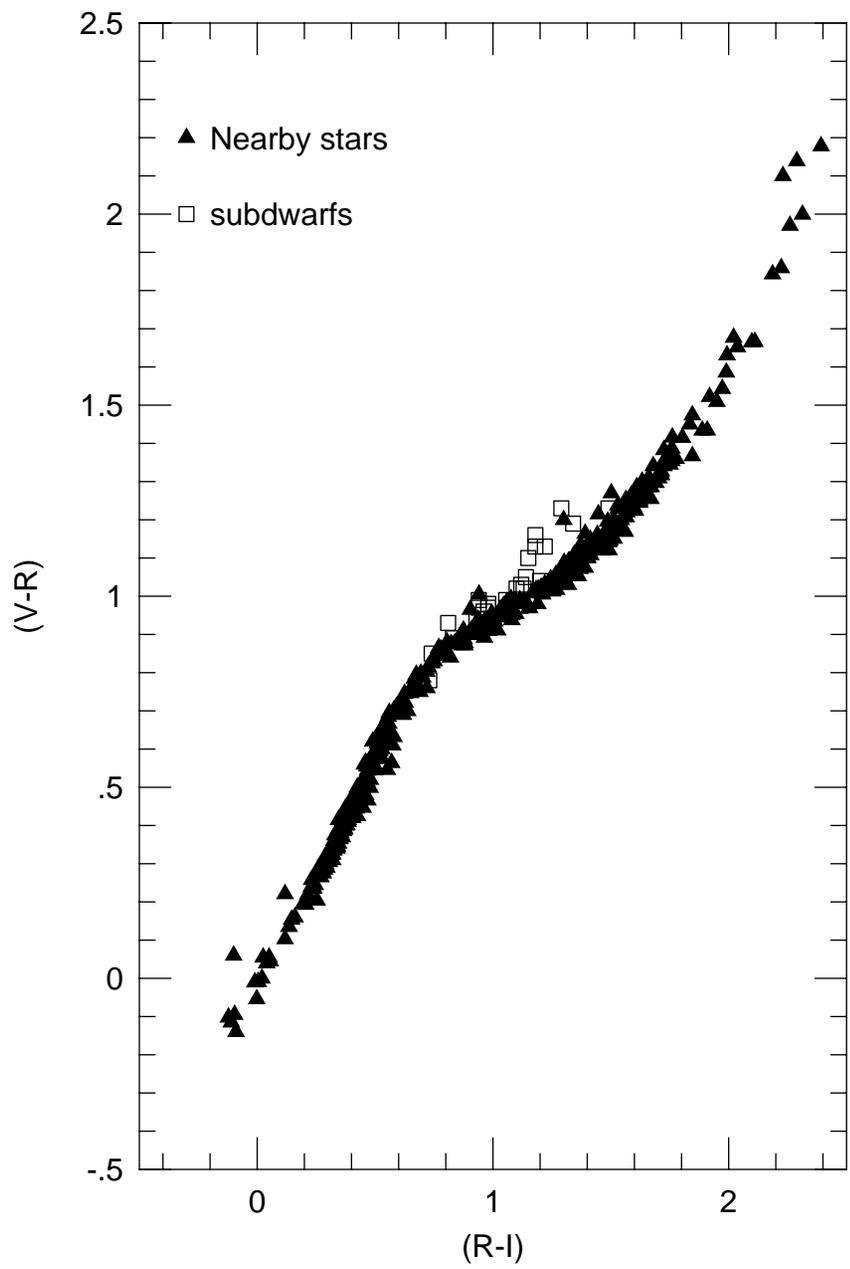
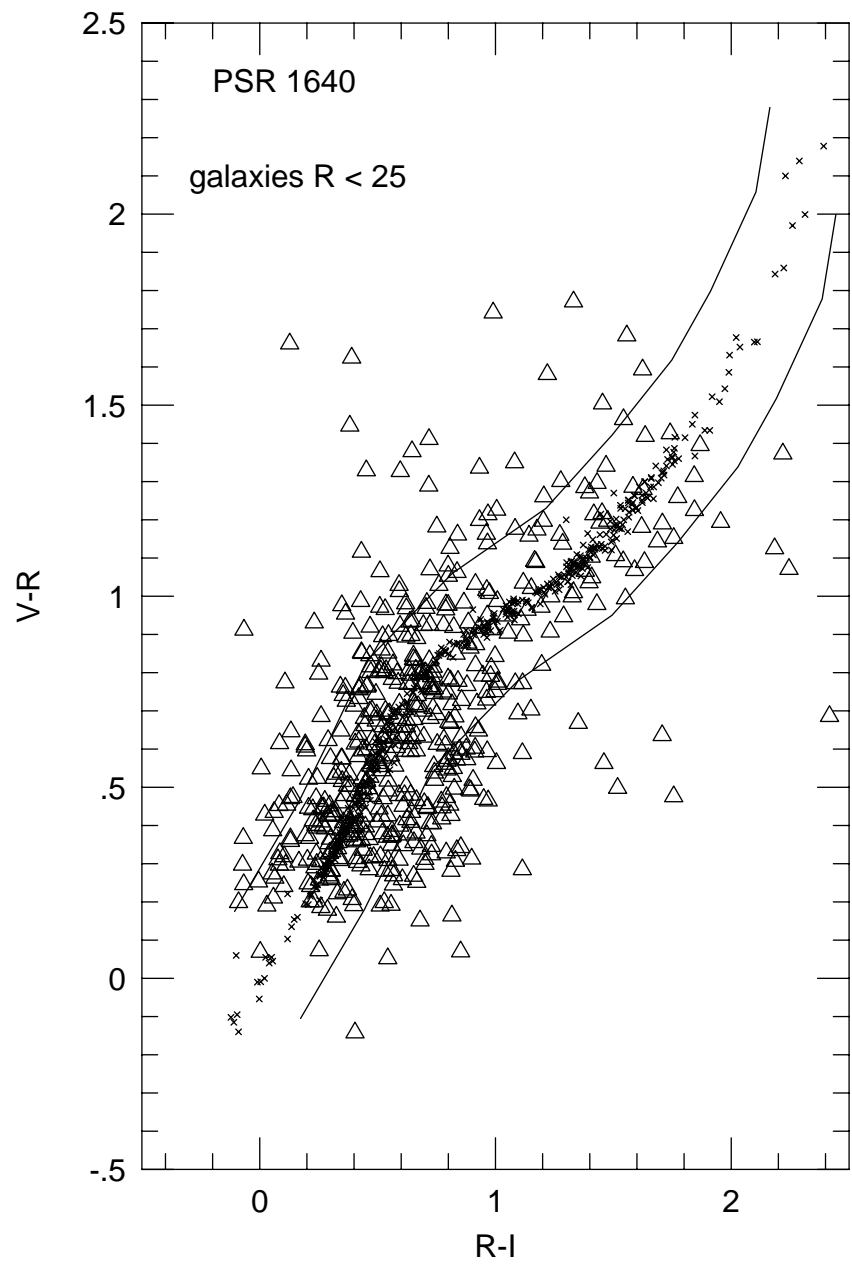
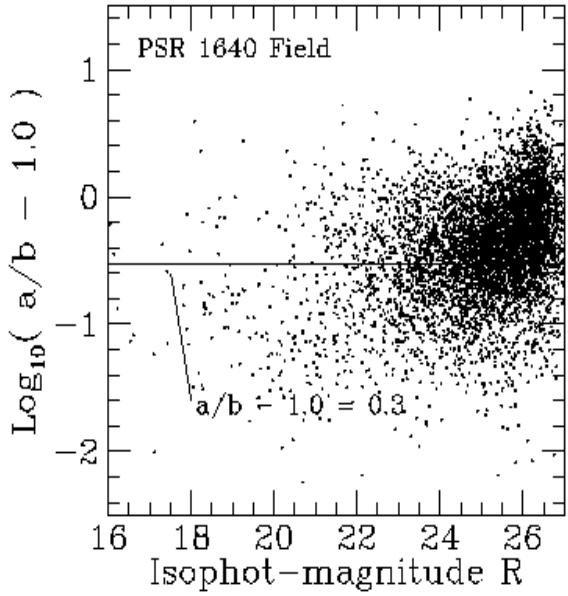
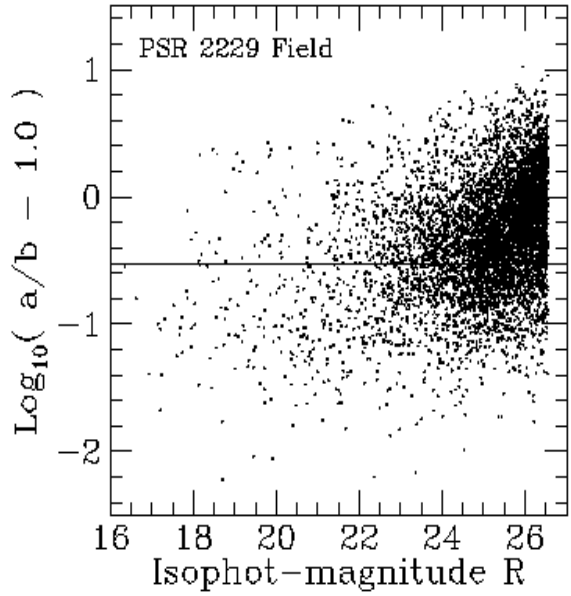
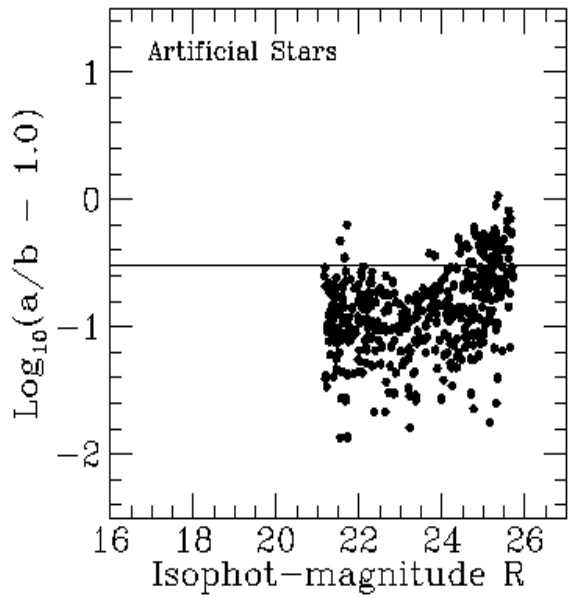
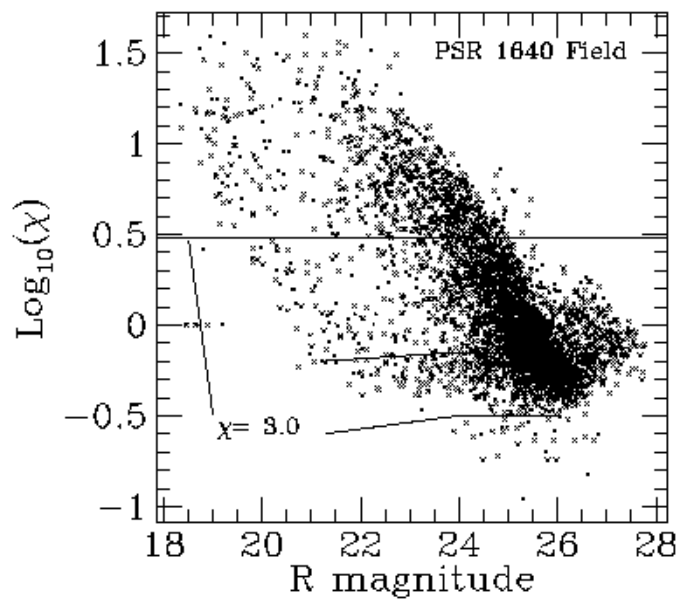
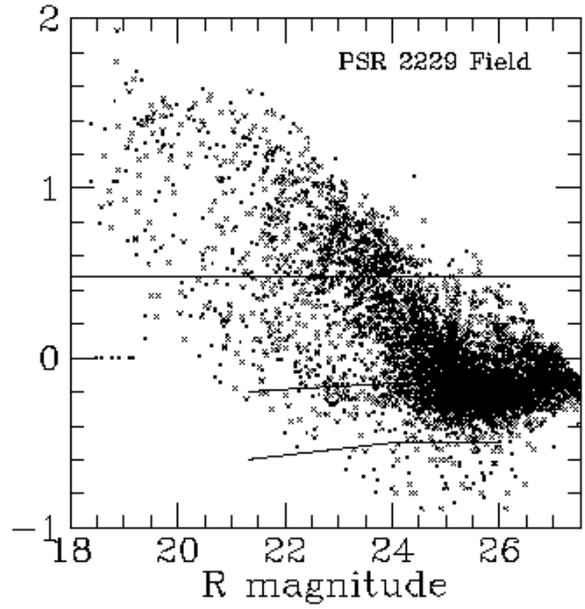
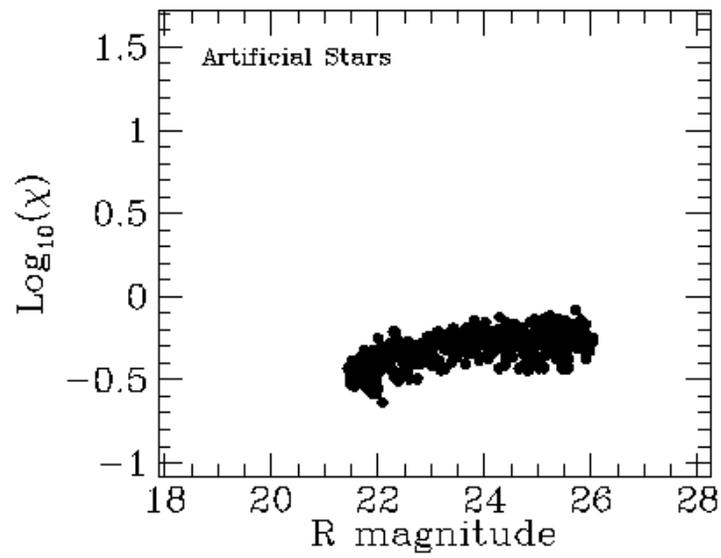
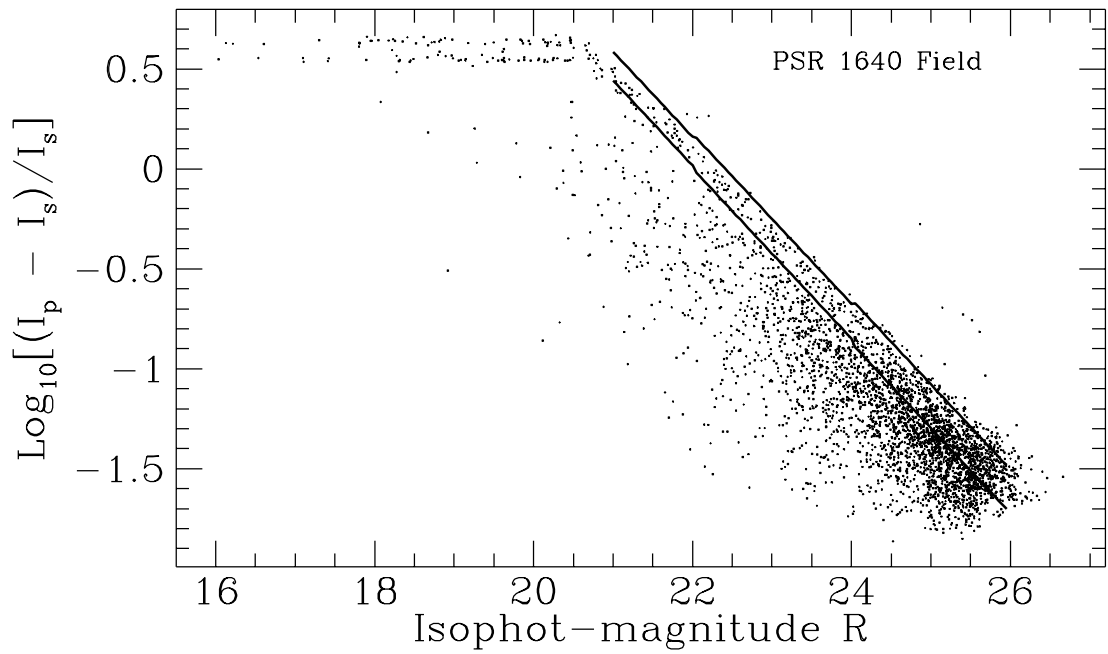
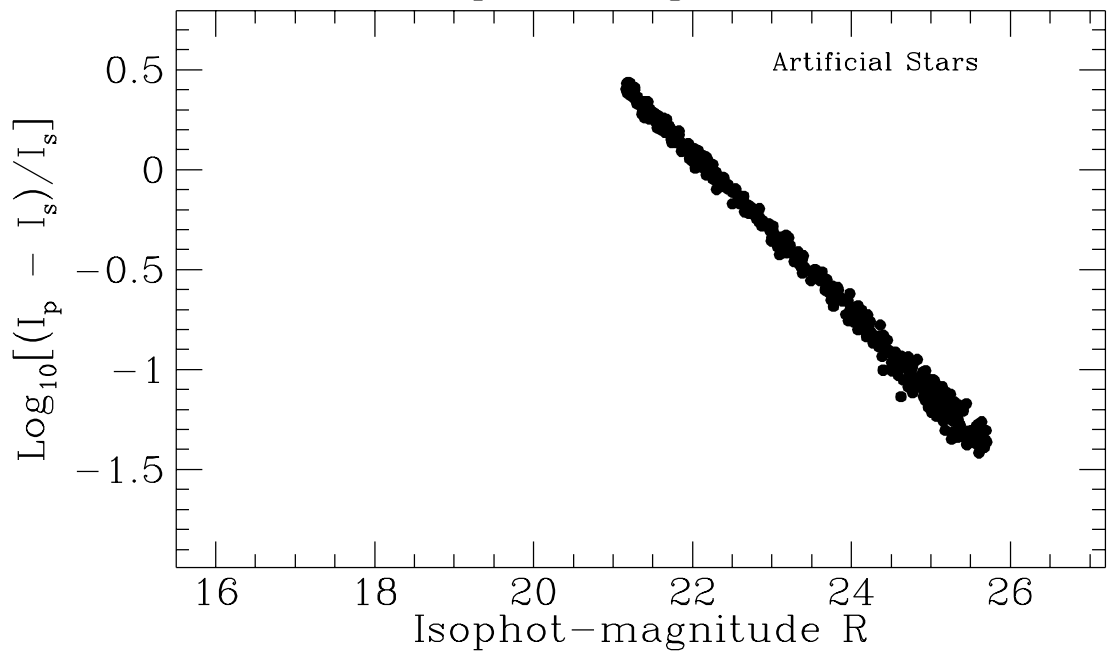
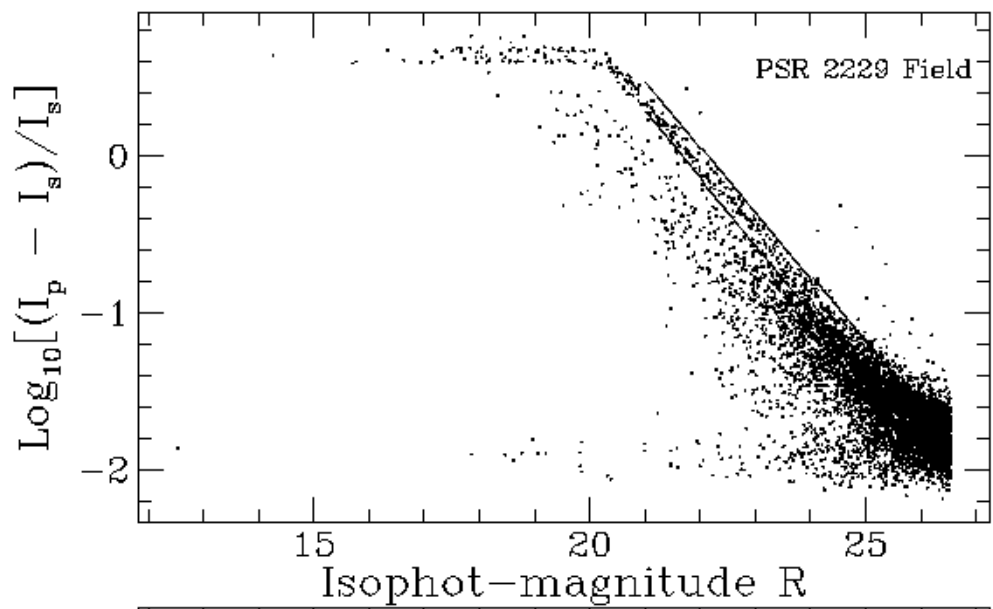
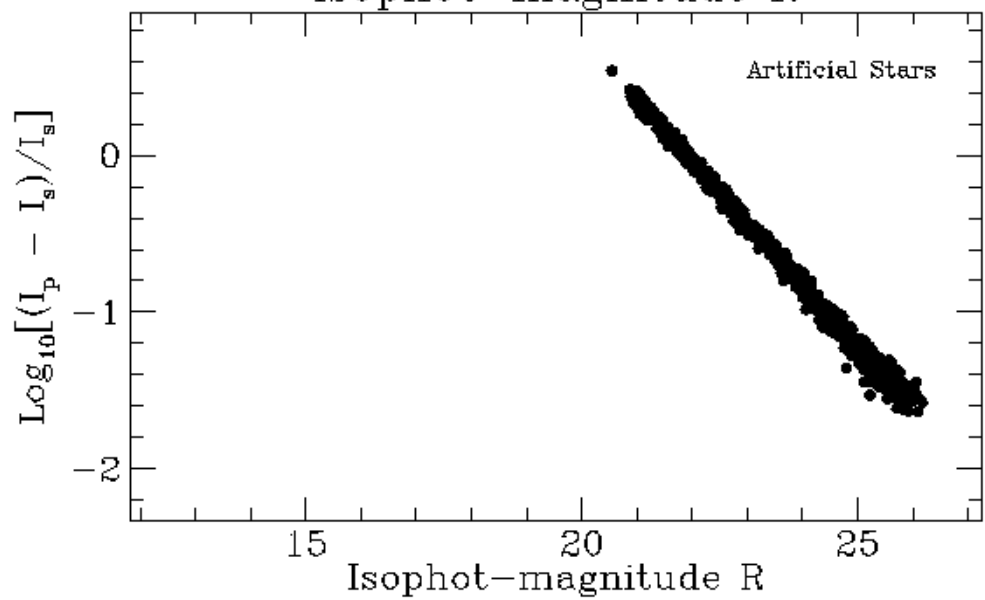
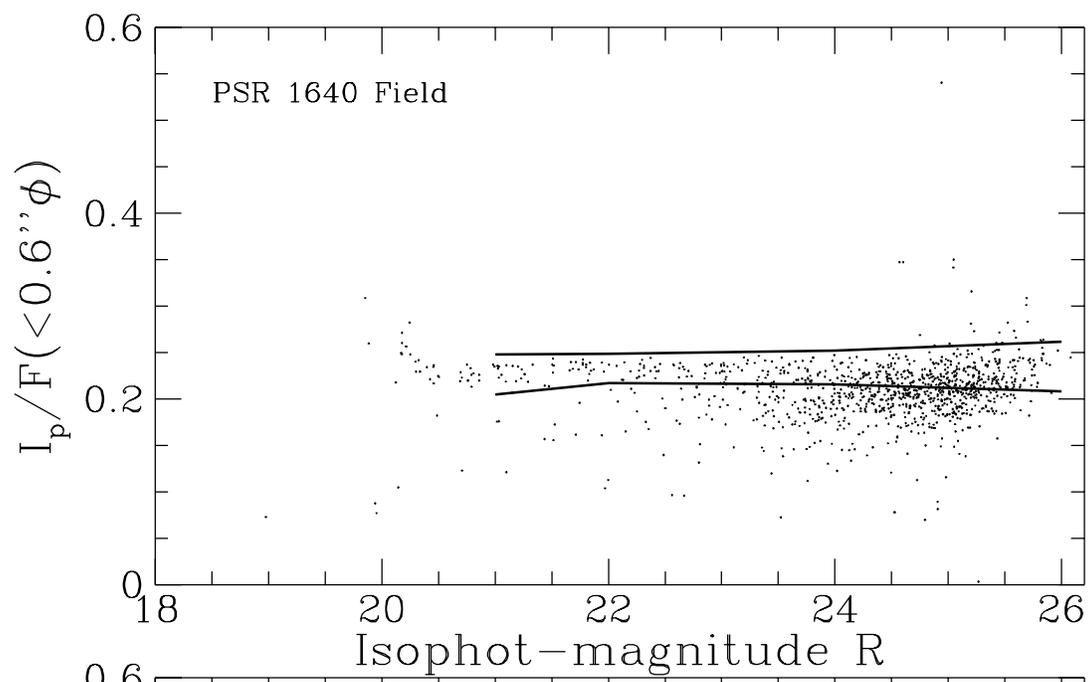
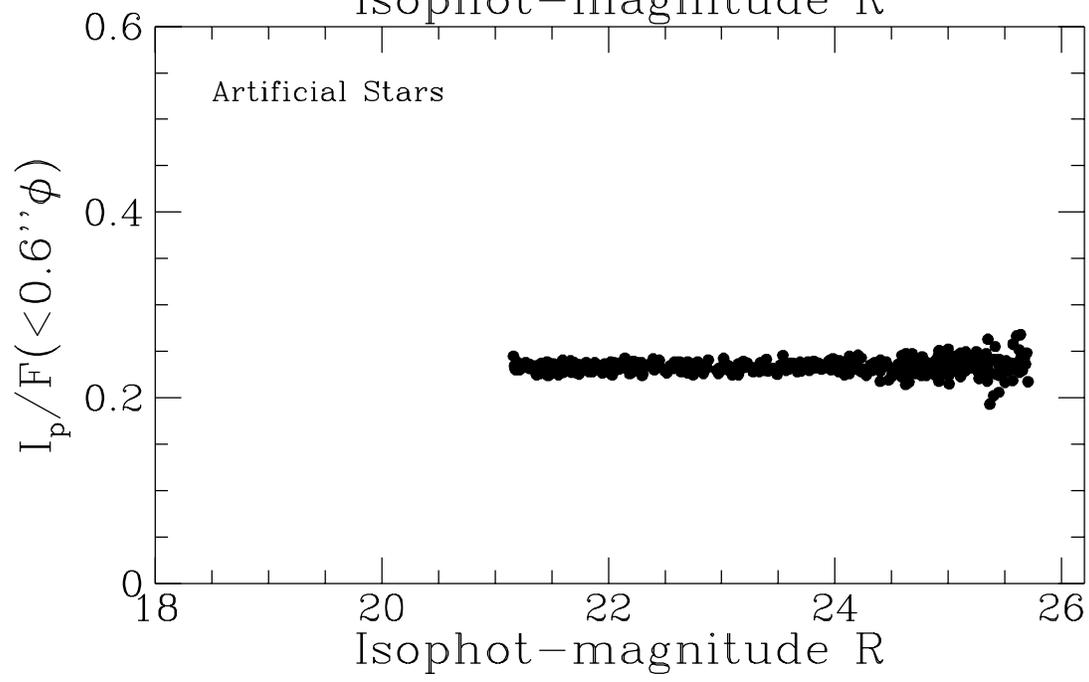
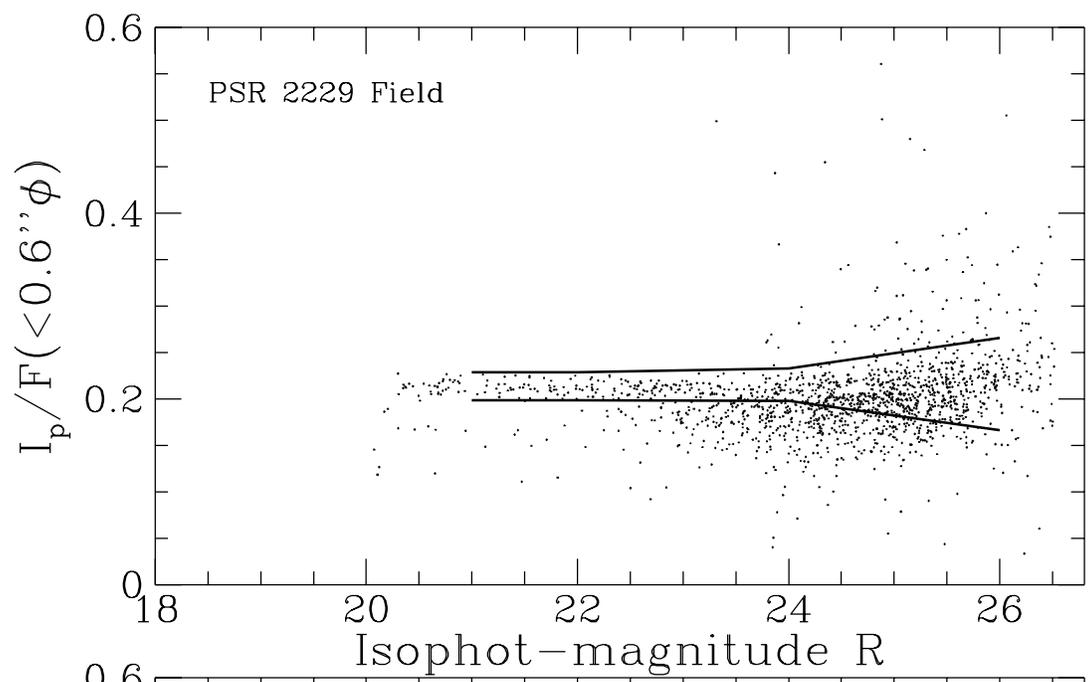
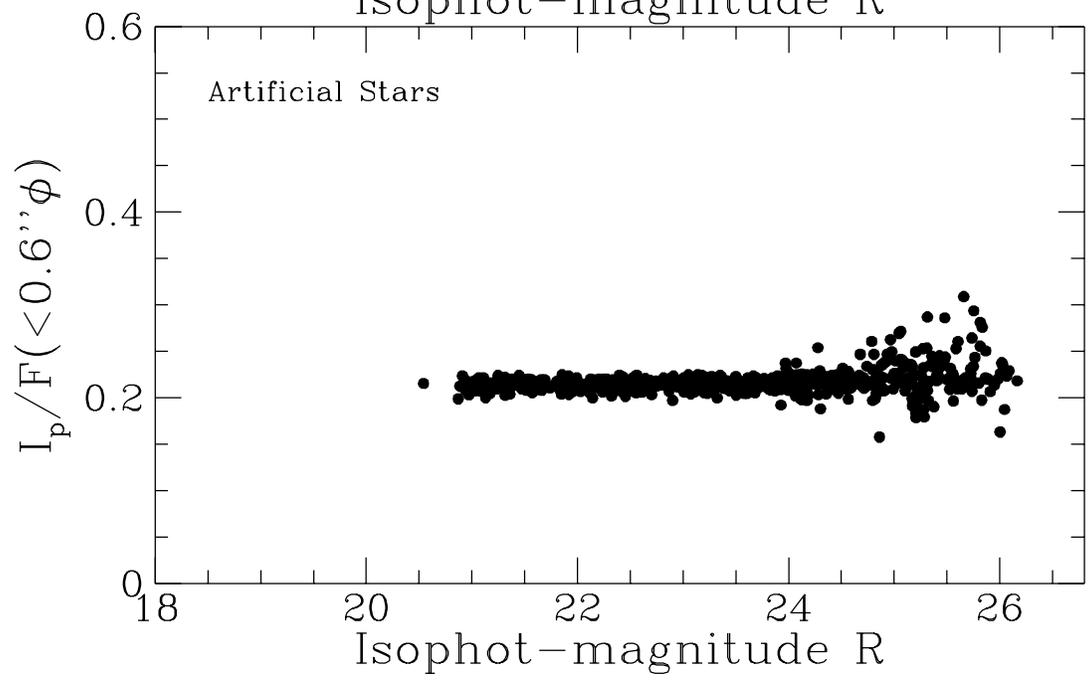
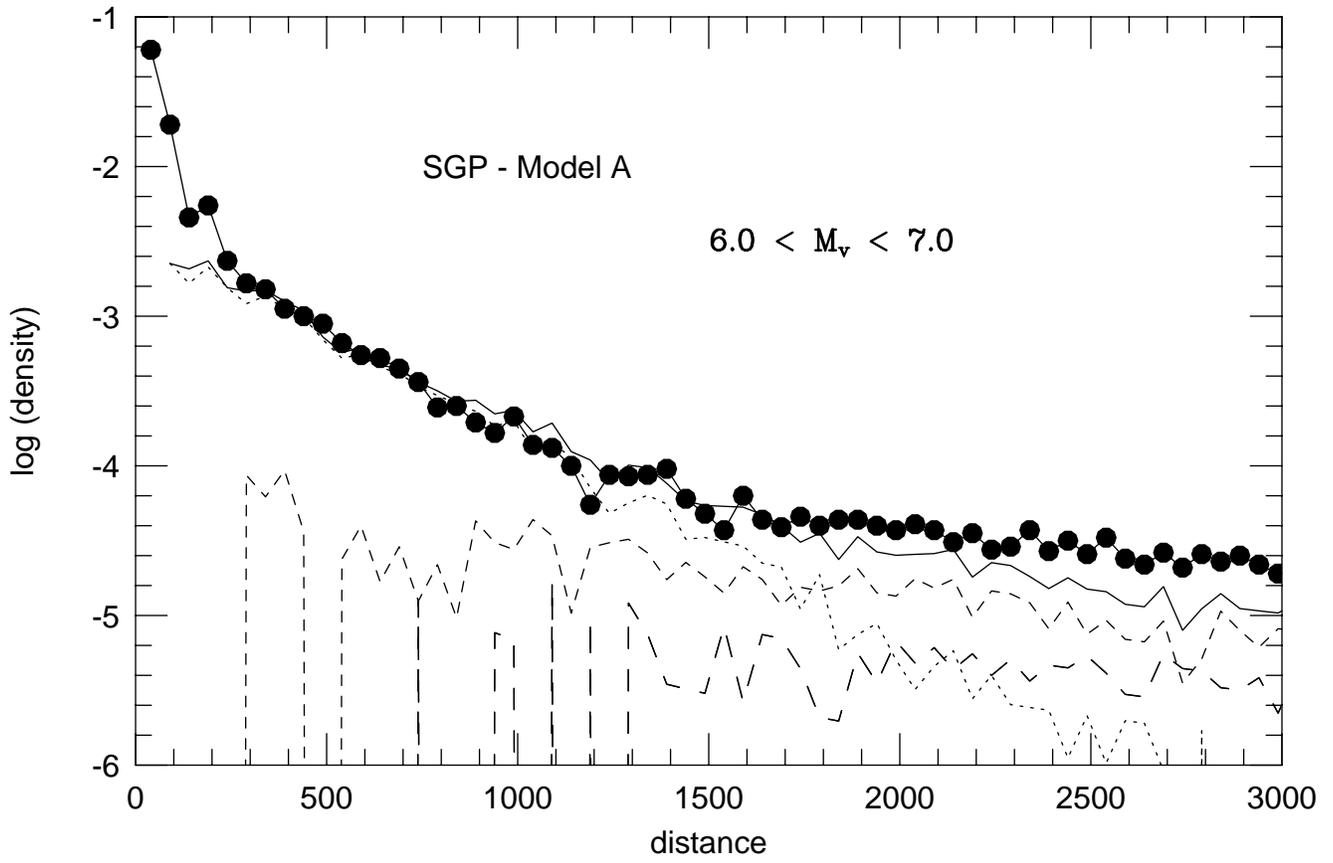
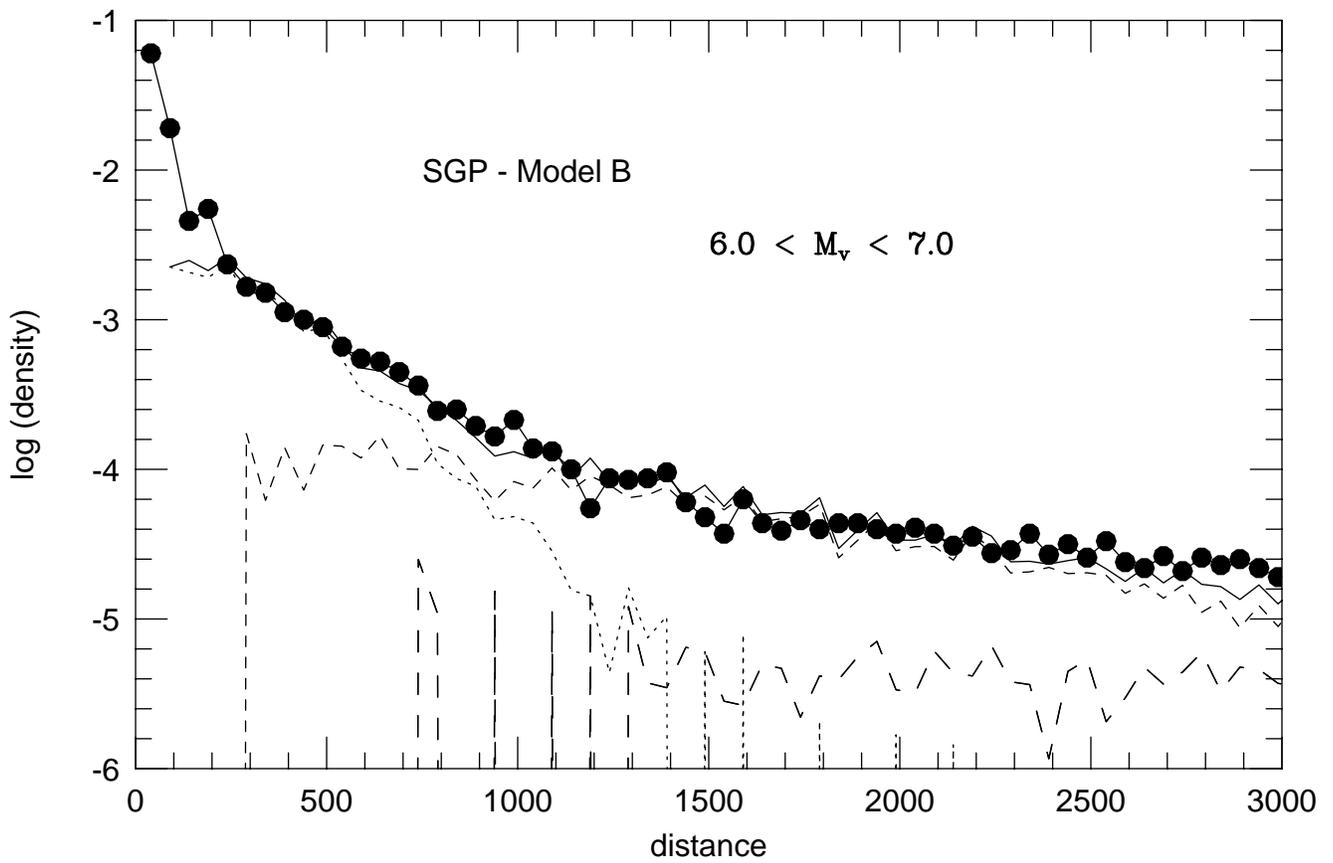
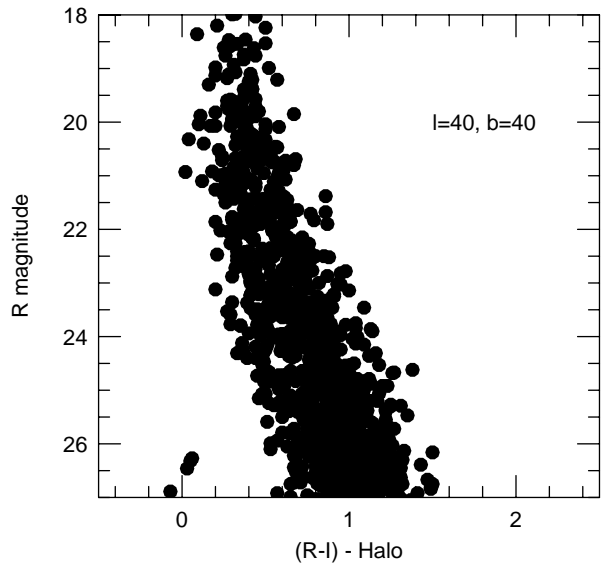
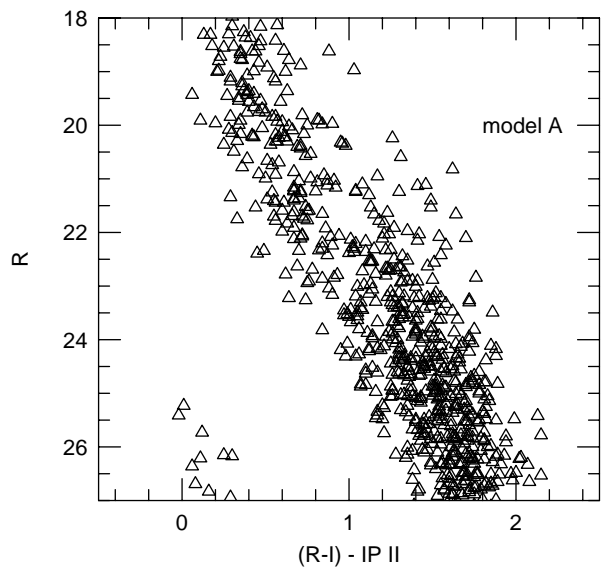
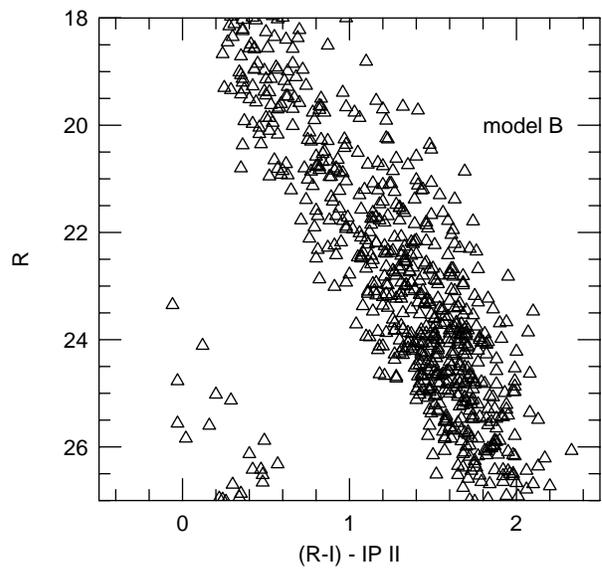
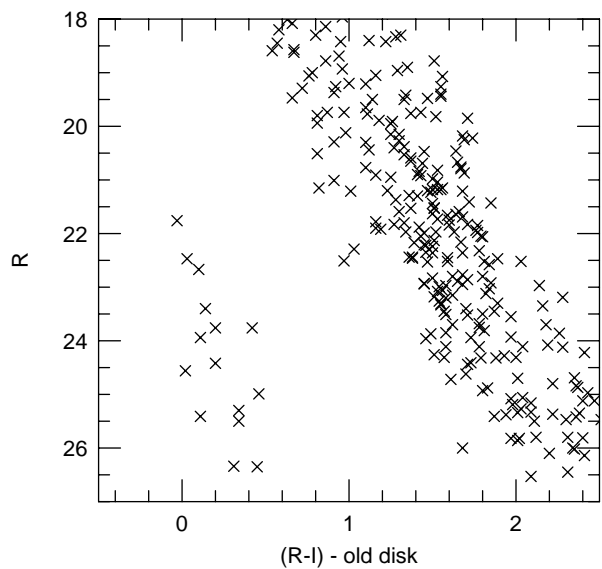
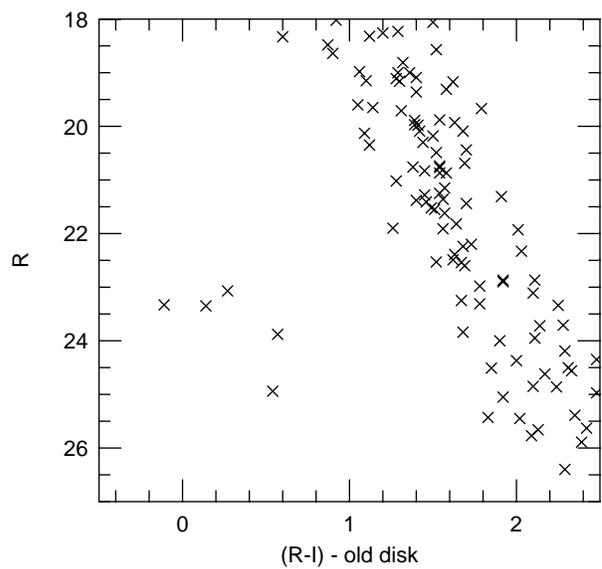
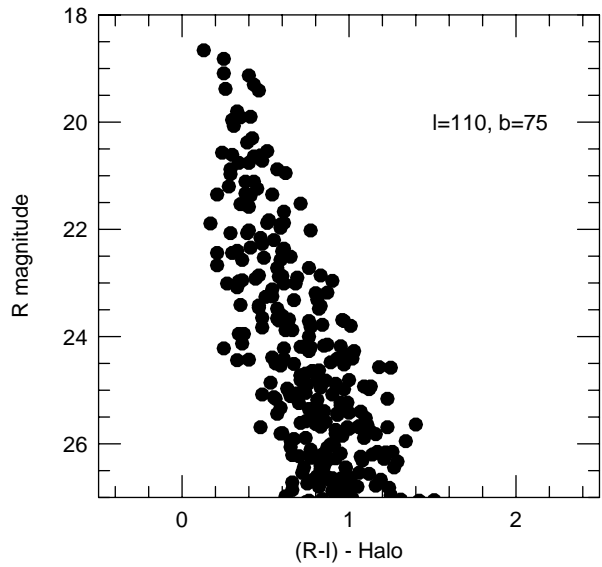
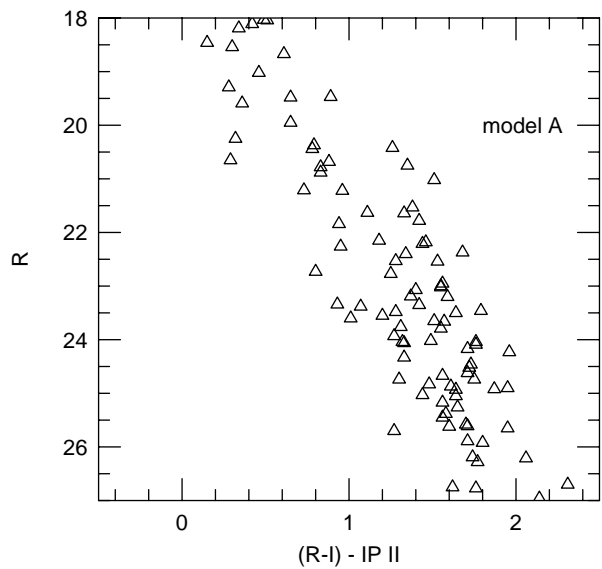
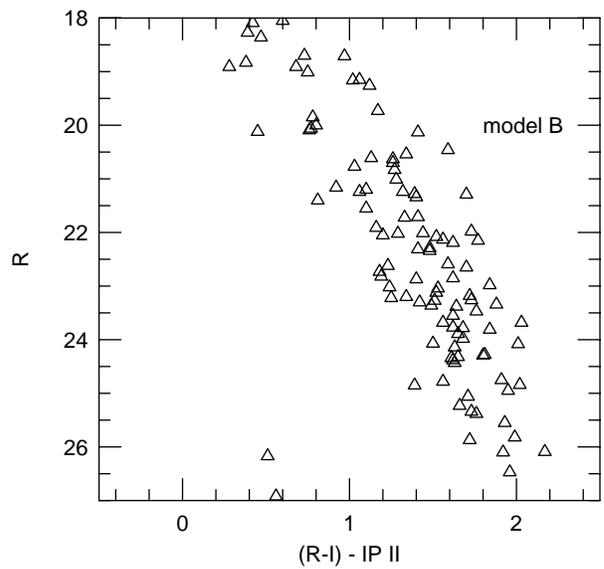
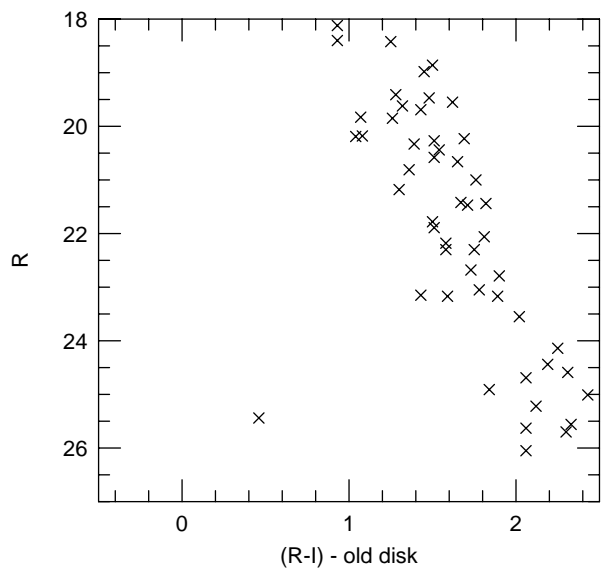
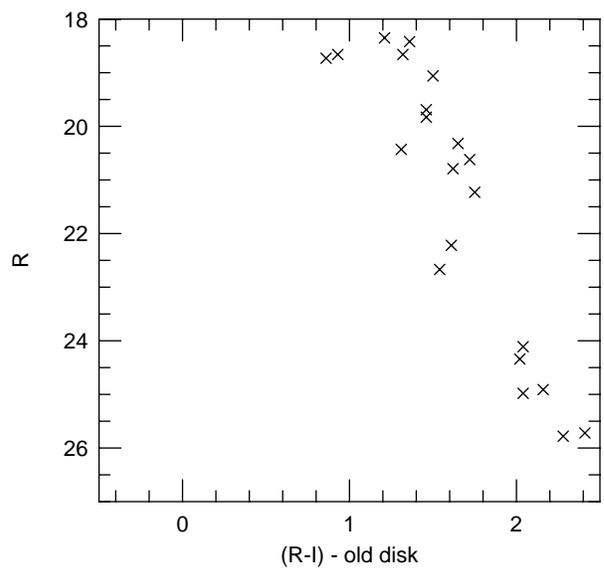
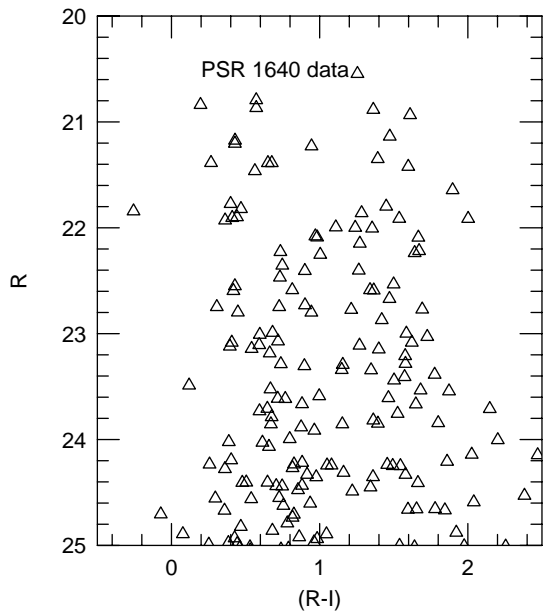
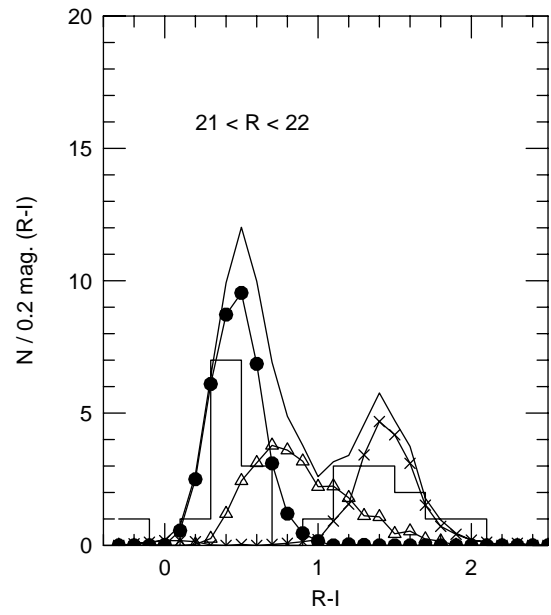
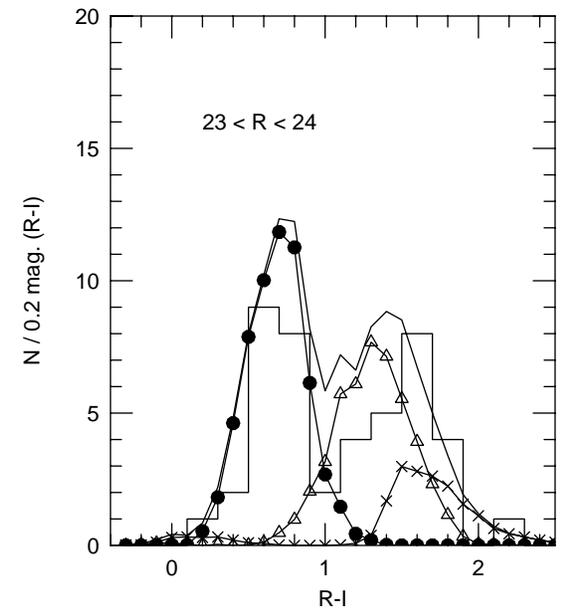
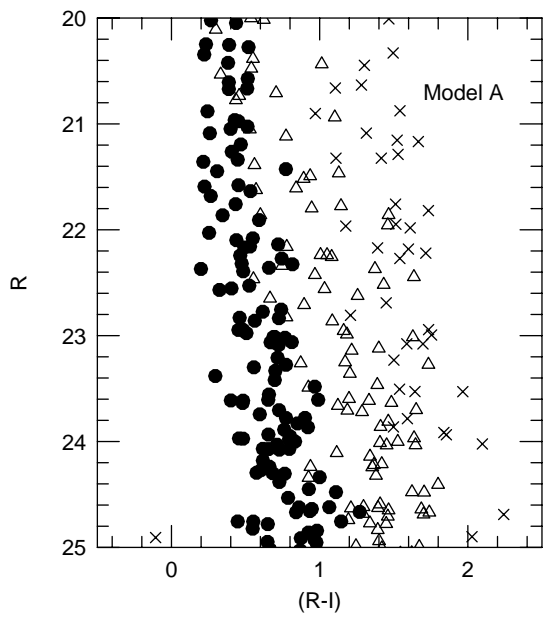
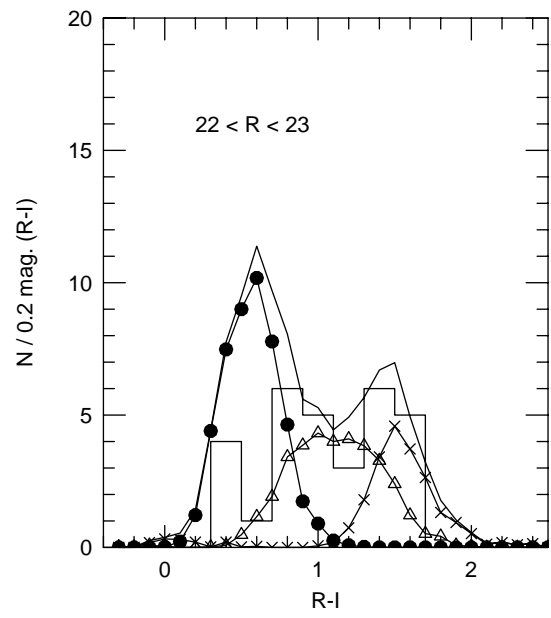
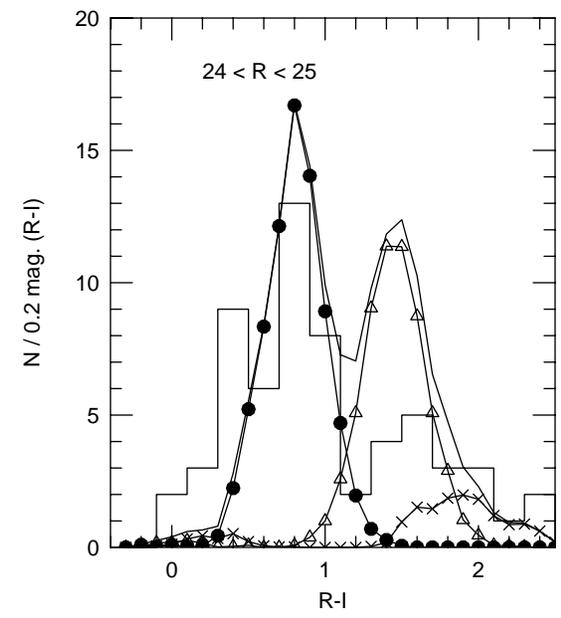
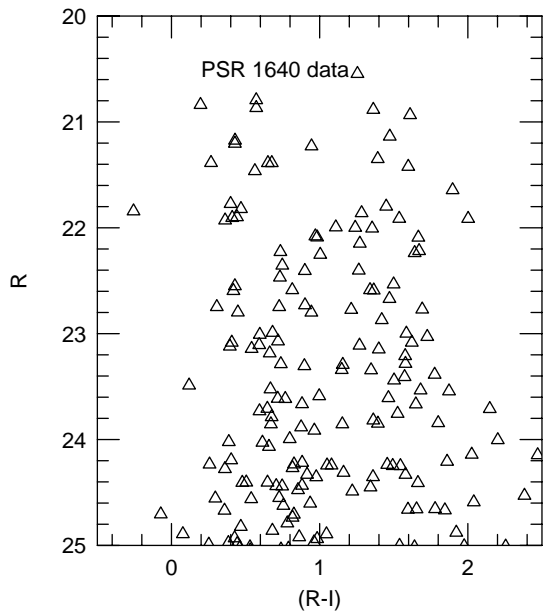
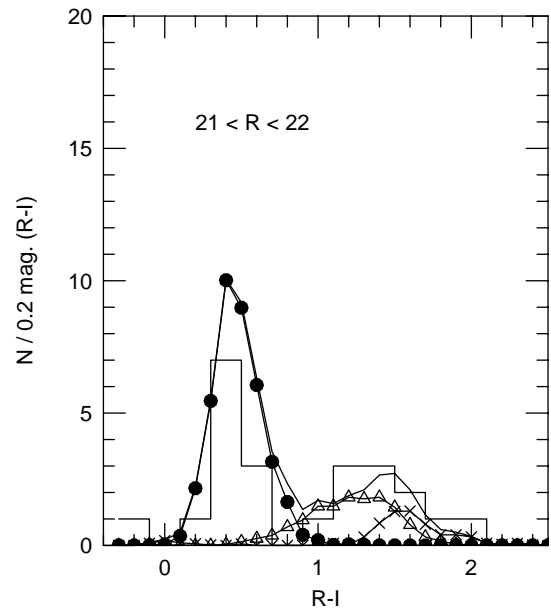
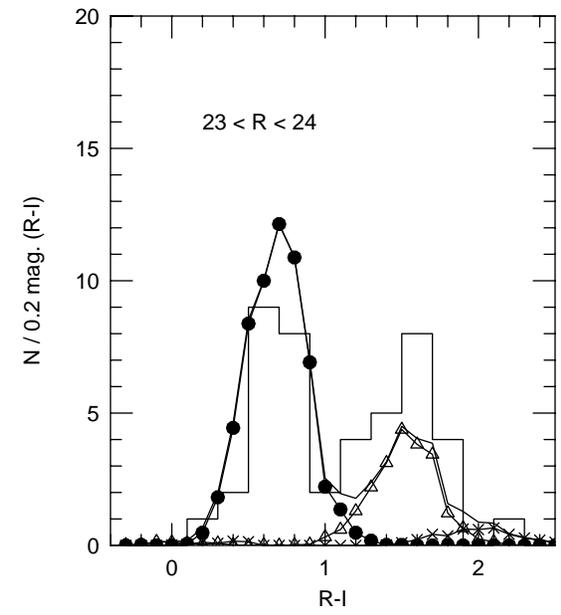
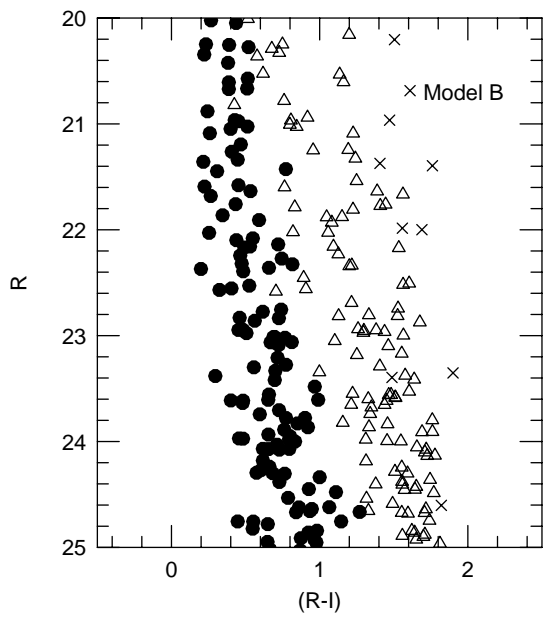
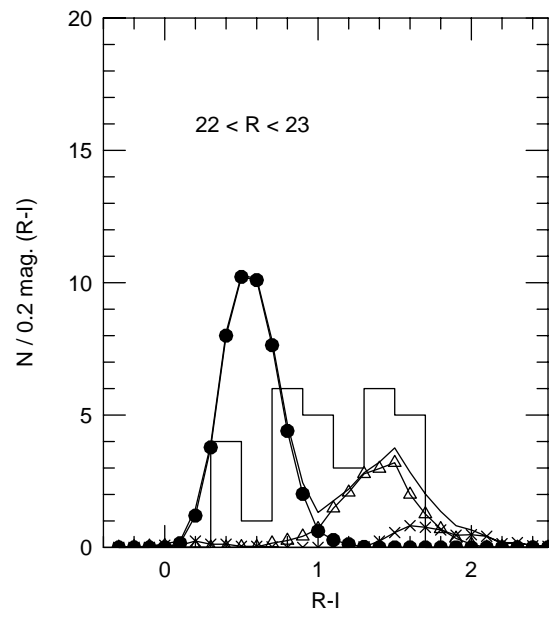
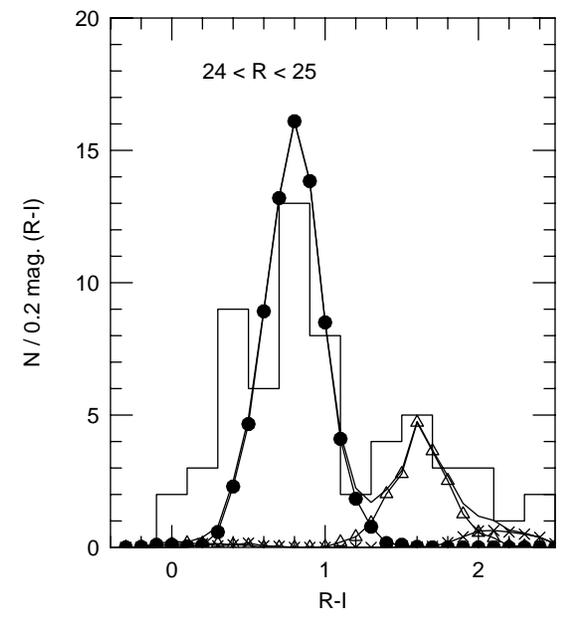
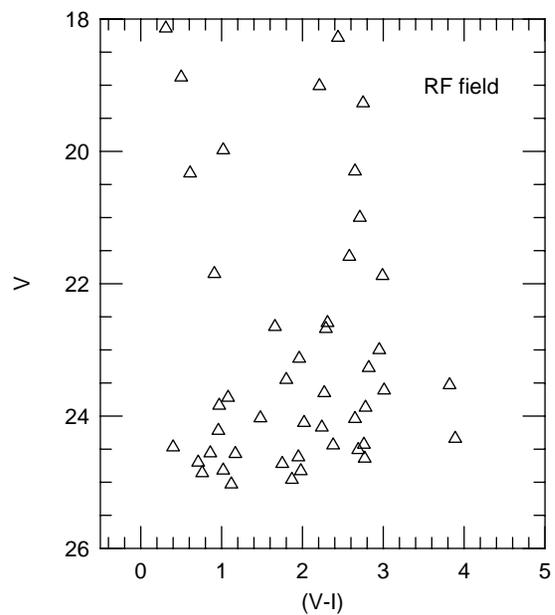
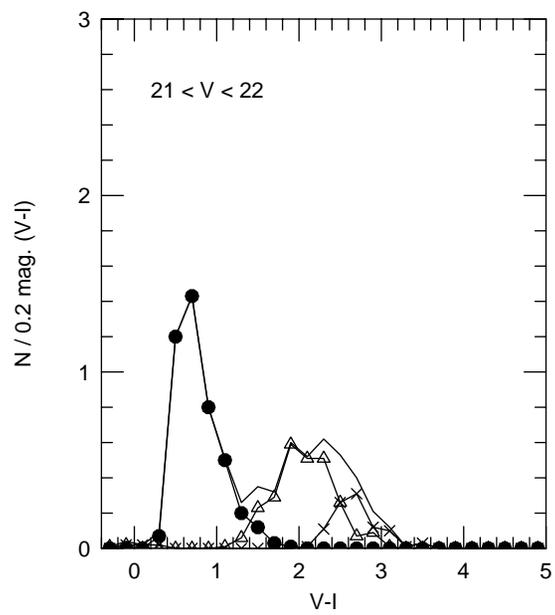
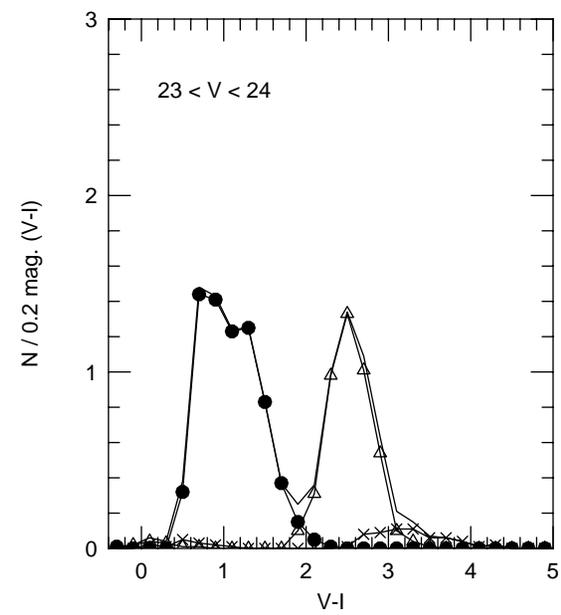
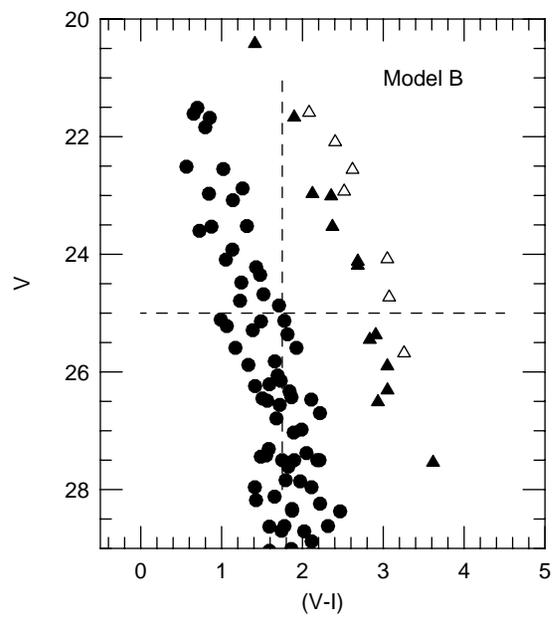
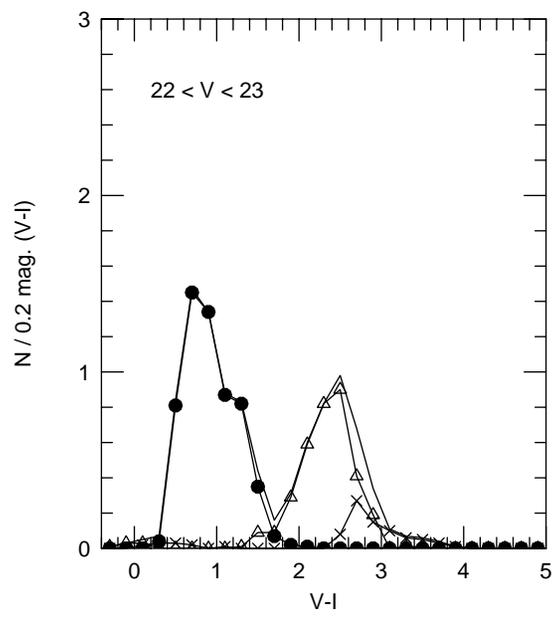
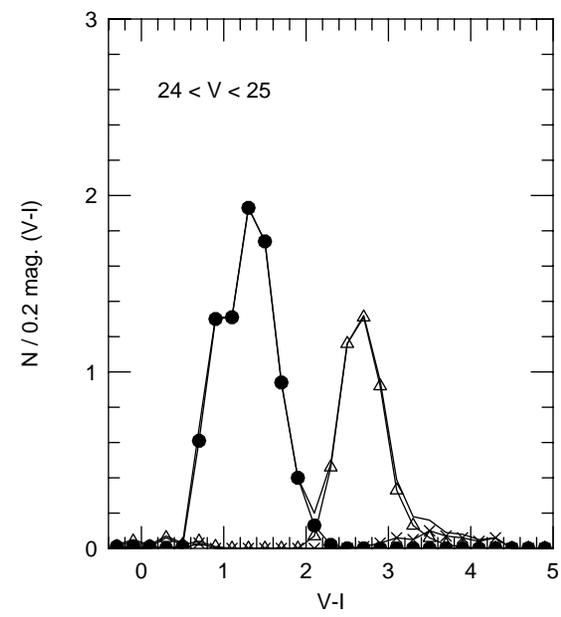
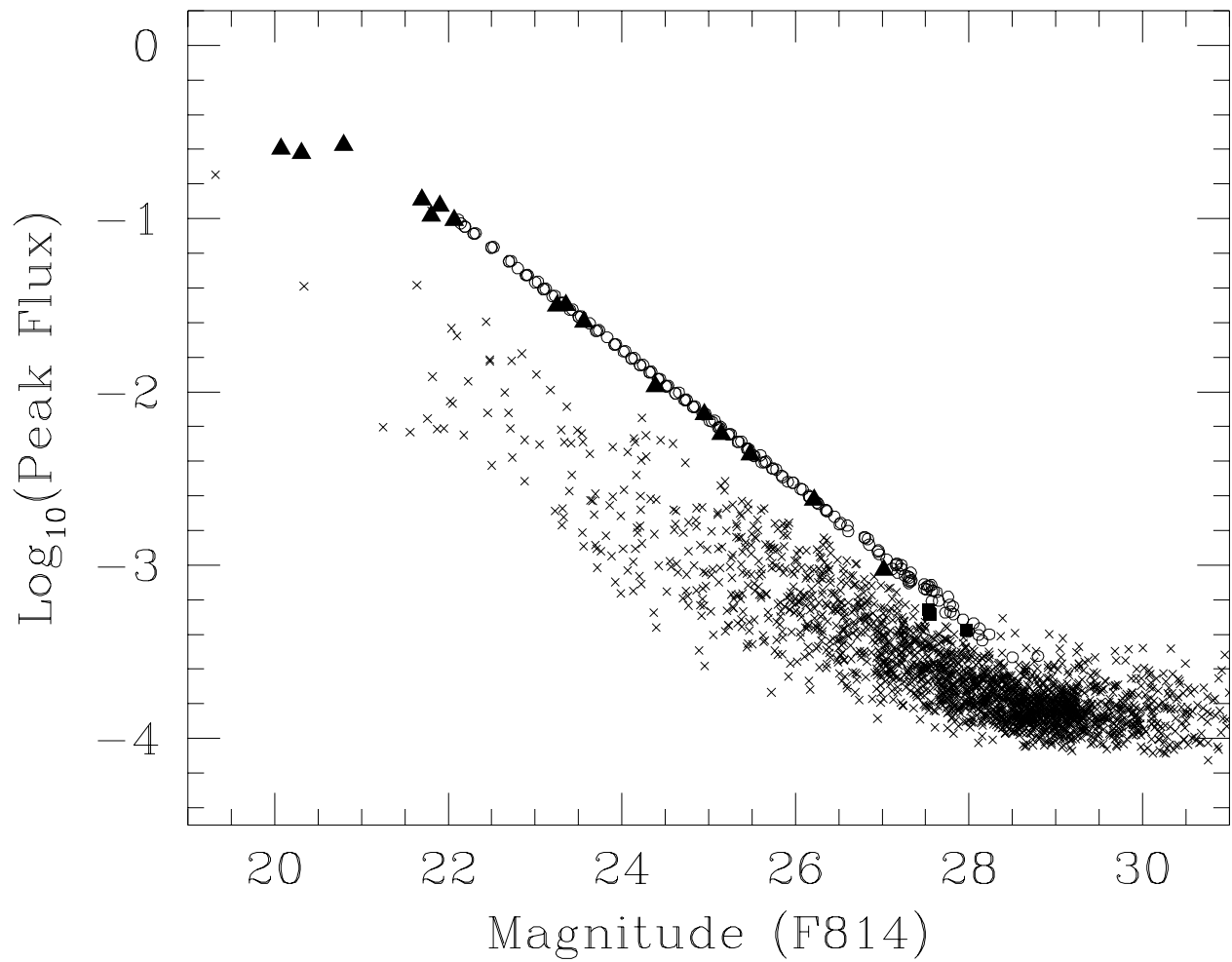
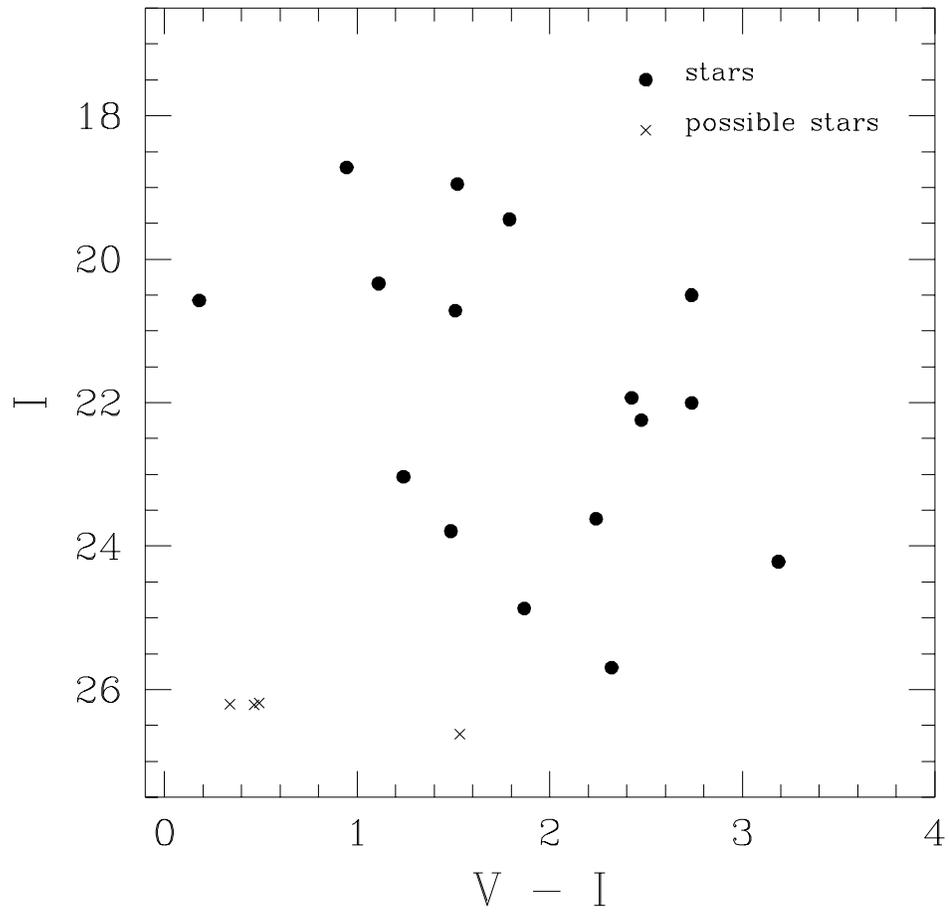
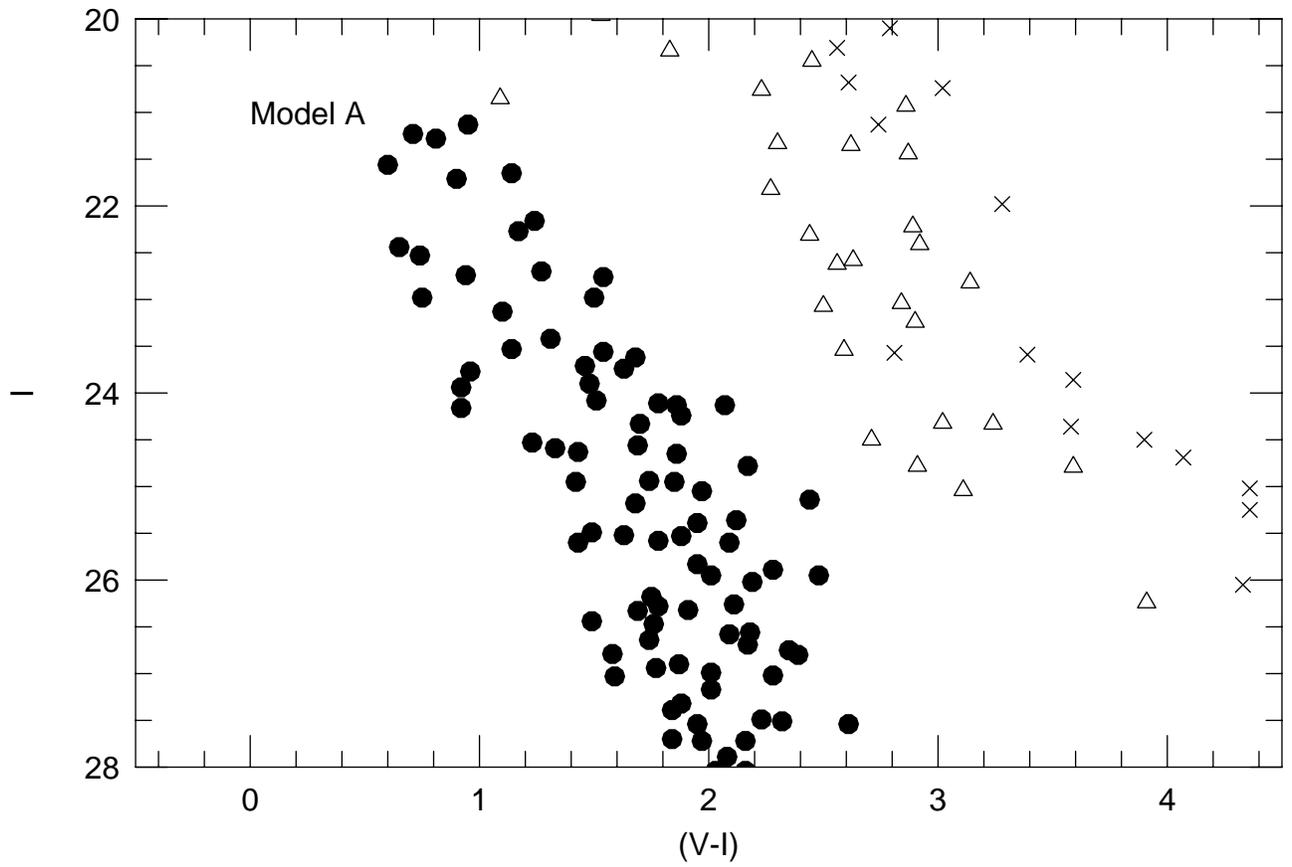
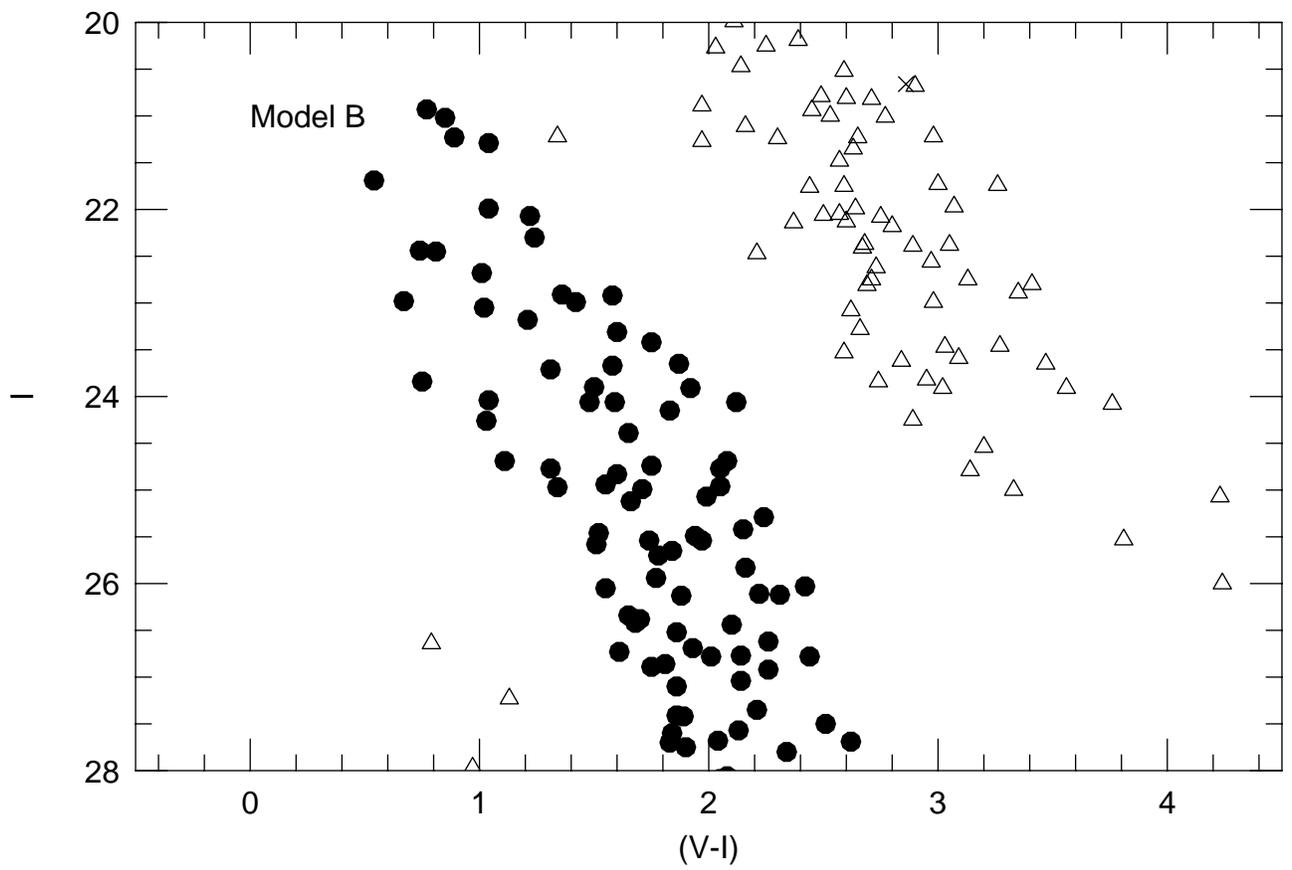
\end